\documentclass[3p,authoryear]{elsarticle}
\usepackage{natbib}
\usepackage{epstopdf}
\usepackage{subfigure,graphicx,float}
\usepackage{amsmath, amsfonts, amssymb}
\usepackage{amsthm}
\usepackage{yhmath}
\usepackage{epsf}
\usepackage{amsfonts}
\usepackage{stmaryrd}
\usepackage{amssymb}
\usepackage{leftidx}
\usepackage{xcolor}
\usepackage{mathtools}
\usepackage{placeins}
\usepackage{booktabs}
\usepackage{enumitem}
\usepackage{caption}
\usepackage{multirow}
\usepackage{stackengine}
\usepackage[utf8]{inputenc}
\usepackage[english]{babel}
\usepackage{bm}
\usepackage{hyperref}
\hypersetup{colorlinks=true,linkcolor=blue,citecolor=blue,urlcolor=blue,linktoc=all}

\DeclareMathAlphabet{\mathsfit}{T1}{\sfdefault}{\mddefault}{\sldefault}
\SetMathAlphabet{\mathsfit}{bold}{T1}{\sfdefault}{\bfdefault}{\sldefault}

\theoremstyle{plain}
\newtheorem{theorem}{Theorem}

\newtheorem{remark}[theorem]{Remark} 
\def\bfS{{\bf S}}
\def\bfT{{\bf T}}

\def\bfF{{\bf F}}

\def\e0{\varepsilon_0}

\def\s0{\sigma_0}

\long\def\symbolfootnote[#1]#2{\begingroup%
\def\thefootnote{\fnsymbol{footnote}}\footnote[#1]{#2}\endgroup}

\makeatletter
\renewcommand\@biblabel[1]{}

\makeatother

\begin{document}
\begin{frontmatter}


\title{Cavitation in elastomers: A review of the evidence against elasticity}

\author{Evan Breedlove}
\ead{elbreedlove@mmm.com}

\author{Chao Chen}
\ead{cchen27@mmm.com}

\author{David Lindeman}
\ead{ddlindeman1@mmm.com}

\author{Oscar Lopez-Pamies}
\ead{pamies@illinois.edu}



\address{3M Corporate Research Laboratory,  St. Paul, MN 55144, USA}

\address{Department of Civil and Environmental Engineering, University of Illinois, Urbana--Champaign, IL 61801, USA}

\begin{abstract}

\vspace{0.2cm}

In spite of the growing body of evidence against it, the elasticity view of the phenomenon of cavitation in elastomers continues to be utilized in numerous studies. In this context, the main objective of this paper is to provide a comprehensive review of the existing evidence that settles that cavitation in elastomers is \emph{not} a purely elastic phenomenon. To that end, a review is first given of the experimental observations of cavitation in elastomers --- gathered since the 1930s until present times --- as well as of its theoretical description as an elastic phenomenon --- whose development started in the 1950s and was substantially completed by the 2010s. The latter is then confronted to the former to pinpoint the reasons why the elastic behavior of elastomers cannot possibly explain the experimental observations. The last part of the paper includes a brief summary of the current view of cavitation as a fracture phenomenon and an outlook for the field in that direction.

\vspace{0.2cm}

\endkeyword

\end{abstract}

\end{frontmatter}

\vspace{0.1cm}

\section{Introduction}

In its original and classical sense\footnote{The term ``cavitation'' has also been recently used to refer to related but different phenomena, which has created some confusion in the literature. We will come back to this important point in Subsection \ref{Sec:Related Problems} below.}, \emph{cavitation} in elastomers refers to the phenomenon of the sudden appearance, or nucleation, of optically visible internal cracks in response to externally applied mechanical forces. The first experimental observations of cavitation reported in the literature date back to the 1930s \citep{Busse38,Yerzley39}. They correspond to a handful of poker-chip experiments on natural and synthetic (neoprene) rubber. These early works went unnoticed until the 1950s, when \cite{GL57,GL59} presented their own poker-chip experiments alongside a first theoretical explanation of the experimental observations. It is the work of \cite{GL57,GL59} that is widely regarded as the origin of the investigations into cavitation in elastomers. In fact, it was \cite{GL57} themselves who coined the term ``cavitation'' by drawing a comparison\footnote{Incidentally, the title of the paper \citep{Bull56} referred to by \cite{GL57} to draw the comparison is `\emph{The tensile strengths of liquids under dynamic loading}'. The significance of this title will become apparent in Subsection \ref{Sec: strength liquids} below.} between their experimental observations and the then-more-mature phenomenon of cavitation in liquids \citep{Rayleigh17,Bull56}.

The famed poker-chip experiments of \cite{GL57,GL59} triggered the emergence of two early schools of thought --- both of them within the framework of continuum mechanics --- to explain the phenomenon of cavitation in elastomers:
\begin{enumerate}

\item{\emph{The elasticity view}. The first school of thought owes its genesis to \cite{GL59} themselves. To explain their experiments, they proposed to view the nucleation of internal cracks in elastomers as the purely elastic growth of their inherent defects.}

\item{\emph{The bottom-up or microscopic fracture view}. The origins of the second school of thought can be traced back to \cite{WS65}. They proposed to view nucleation of internal cracks in elastomers as the growth of their inherent defects \emph{not} by elastic deformation but by the creation of new surface --- that is, by fracture --- as dictated by the \cite{Griffith21} competition between the bulk elastic energy and the surface fracture energy of the elastomer.}

\end{enumerate}

Presumably because of the alluring simplicity of its most elementary mathematical model --- that is, the radially symmetric elastic deformation of a single spherical cavity of  infinitesimally small size embedded in a homogeneous elastic medium of finite size under remotely applied pressure (see Subsection \ref{Sec: The first result} below) --- together with its apparent agreement with some of the experimental observations, the elasticity view is the one that has been most frequently employed \citep{Horgan95,Fond01,LP11a}. Despite delivering similar apparent agreement with some of the same experimental observations, the microscopic fracture view has not received nearly as much attention \citep{Gent90,GentWang91}. This is surely because of the technical difficulties associated with having to model explicitly at the same time the presence of the inherent defects in elastomers, the creation of new surface at the small length scale of those defects, and the growth/transition of that new surface to a macroscopic length scale.

\begin{remark}
\emph{Here, it is important to recall and emphasize that, as continuum theories, both the elasticity and the microscopic fracture views can only model the inherent defects in elastomers within the bounds afforded by the idealization of matter as a continuum. In such an idealization, a cavity or crack of infinitesimally small size refers to a cavity or crack whose size can be considered to be vanishingly small relative to any other continuum length in the problem at hand, but one that is still larger than the characteristic molecular length scale of the elastomer at hand.}
\end{remark}

Despite the very many investigations of cavitation as a purely elastic phenomenon since the 1950s, it was only in 2015 that a direct comparison\footnote{See, however, the work of \cite{Abeyaratne89} for an earlier preliminary comparison.} between the elasticity view and the founding poker-chip experiments of \cite{GL59} was reported in the literature by \cite{LRLP15}. That direct comparison indicated that, while the nonlinear elasticity of the elastomers is important, cavitation as reported by \cite{GL59} is first and foremost a fracture process.

Since the analysis of \cite{LRLP15}, a number of newly designed experiments, making use of modern diagnostics tools, and their corresponding analyses \citep{Poulain17,Poulain18,KFLP18,KRLP18,GuoRavi23,KKLP23} have repeatedly corroborated that cavitation in elastomers is indeed a fracture process and \emph{not} a purely elastic one. What is more, these works have prompted the emergence of a new school of thought --- also within the framework of continuum mechanics --- to explain the phenomenon of cavitation in elastomers:
\begin{enumerate}\setcounter{enumi}{2}

\item{\emph{The top-down or macroscopic fracture view}. The third school of thought finds its origin in the work of \citet*{KFLP18}. These authors proposed to view the nucleation of internal cracks in elastomers as a fracture phenomenon from a top-down or macroscopic perspective. Accordingly, the inherent defects in an elastomer --- in stark contrast to the two previous schools of thought ---  are \emph{not} modeled explicitly, instead, it is their macroscopic manifestation that is modeled explicitly, that is, the \emph{strength} of the elastomer.}

\end{enumerate}

An additional fundamental difference between the macroscopic fracture view initiated by \citet*{KFLP18} and the earlier elasticity and microscopic fracture views initiated by \cite{GL59} and \cite{WS65} is that it is centered around a \emph{complete} theory of nucleation and propagation of fracture, as opposed to being solely focused on describing the sudden appearance of internal cracks without regard of the ensuing behavior. A string \citep{KFLP18,KRLP18,KLP20,KLP21,KKLP23} of recent direct comparisons with both new and classical experiments has shown that the approach initiated by \citet*{KFLP18} is indeed capable of accurately describing the entire range of experimental observations of cavitation found in the literature; Subsection \ref{Sec:Fracture Comparisons} below presents a few representative results.

In spite of the growing body of evidence against it since the analysis of \cite{LRLP15}, the elasticity view of cavitation continues to be widely utilized. In this context, the main objective of this paper is to provide a comprehensive review of the existing evidence that conclusively establishes that cavitation in elastomers cannot be explained as a purely elastic phenomenon.

We begin in Sections \ref{Sec: Experiments} and \ref{Sec: Elasticity Theory} by providing reviews of the current experimental knowledge of cavitation in elastomers and of the elasticity view of the phenomenon. In Section \ref{Sec: Theory vs Experiments}, we then confront the latter to the former and thereby pinpoint the reasons why cavitation in elastomers is not a purely elastic phenomenon. There are essentially two reasons:
\begin{enumerate}[label=\roman*.]

\item{The elasticity of elastomers is non-Gaussian.}

\item{The location where cavitation occurs does \emph{not} correlate with the location where defects would grow elastically.}

\end{enumerate}
We discuss these two reasons one at a time in Subsections \ref{Sec: non-Gaussian} and \ref{Sec: Location}, respectively. We conclude in Section \ref{Sec:Final comments} by providing a brief summary of the current view of cavitation as a macroscopic fracture phenomenon and an outlook for the field in that direction. This includes a discussion of related emerging problems not just in elastomers but in other soft matter systems.

\section{A summary of direct and indirect experimental observations of cavitation}\label{Sec: Experiments}

Experiments have shown that cavitation in elastomers typically occurs at material points where the state of stress is in the first octant in the space of principal Cauchy stresses $(t_1,t_2,t_3)$, that is, where $t_1,t_2,t_3>0$. Such states of \emph{all-tensile triaxial stress} are difficult to realize directly with a testing machine. However, due to the inherent near incompressibility of elastomers, such stress states naturally develop around the interface of an elastomer with a stiff domain when pulling the two apart. In fact, nearly all the experimental observations of cavitation that have been reported in the literature make use, in one way or another, of such a configuration.

In this section, we provide a summary of all of these experimental observations, classified in terms of the geometry of the stiff domain around which cavitation occurs. Within each type of geometry, so as to also provide a modicum of historical perspective, we present the observations in chronological order, as they have been reported in the literature since the 1930s until present times.

\subsection{Cavitation in poker-chip experiments}\label{Sec: Poker-Chip}

By far, the experiment that has been utilized the most to study cavitation in elastomers is the poker-chip experiment \citep{Busse38,Yerzley39,GL57,GL59,Lindsey67,Kakavas91,Dorfmann03,Creton10,Bayraktar08,Drass18,Euchler20,GuoRavi23}. As schematically depicted in Fig. 1, this experiment consists in firmly bonding a thin disk of an elastomer to stiff fixtures and then pulling the fixtures apart in a testing machine.
\begin{figure}[h!]
\centering
\centering\includegraphics[width=0.9\linewidth]{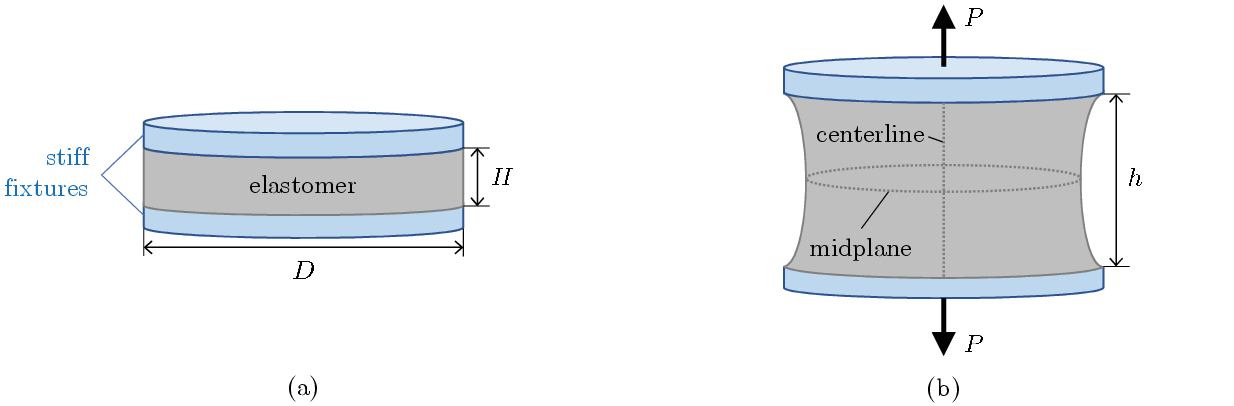}
\caption{\small Schematic of the poker-chip experimental setup in (a) the initial configuration and in (b) a deformed configuration at an applied deformation $h$ and corresponding tensile force $P$. Relative to its initial thickness $H$, the initial diameter $D$ of the elastomer disk --- the ``poker chip'' --- is typically in the range $D/H\in[2,50]$. For reference in the text, the midplane and centerline of the specimen are marked in part (b).}\label{Fig1}
\end{figure}

From the early poker-chip experiments carried out by \cite{Busse38}, only the relation between the normalized global force $S=4P/(\pi D^2)$ and the normalized global deformation $\lambda=h/H$ measured from three specimens with diameter-to-thickness ratios $D/H=4,15,23$, all made of the same natural rubber, was reported, alongside several post-mortem images of the midplane of the specimens cut open showing the presence of multiple cracks. Thus, while the evidence presented by \cite{Busse38} made clear that cracks had appeared in the interior of the natural rubber, it did not shed light on where and when those cracks nucleated, on whether they had propagated prior to the instance at which the images of the midplanes were taken, nor on whether there were additional cracks elsewhere in the specimens that were not visible at their midplanes.

Similar to \cite{Busse38}, \cite{GL59} also presented results for the normalized global force-deformation ($S$ \emph{vs}. $\lambda$) relation and post-mortem images of the midplanes of the specimens. Critically, in contrast to \cite{Busse38}, \cite{GL59} made use of specimens with numerous (in excess of twenty) diameter-to-thickness ratios $D/H$ spanning the range $D/H\in[2,50]$ and carried out experiments on eight different elastomers with a variety of different elastic behaviors. Specifically, they carried out experiments on seven different vulcanizates of natural rubber, four unfilled and the other three filled with different volume fractions of carbon black, as well as on one type of styrene-butadiene rubber (SBR). One of the unfilled vulcanizates (labeled \texttt{G} in their paper) was transparent, which allowed them to observe \emph{in situ} the appearance of cracks within the specimen during the loading process. The wealth of experimental observations generated by \cite{GL59} can be summarized as follows:

\begin{itemize}

\item{\emph{The what}. The primary result of \cite{GL59} is that they observed, both \emph{in situ} and \emph{post mortem}, the appearance of internal cracks in the elastomers that they tested at some critical separation between the fixtures. Crucially, their observations pertain to the sudden appearance of \emph{optically visible} cracks. Since the spatial resolution of an optical microscope is about 1 $\mu$m, a consistent and unambiguous definition is then that cavitation is the sudden appearance of internal cracks of size $\geq 1\mu$m.}

\item{\emph{The where}. For the thinner specimens (roughly $D/H>20$), cavitation occurs around the midplane. Precisely, according to the transparent vulcanizate \texttt{G}, the sequence of events is such that a crack (or a few cracks) first appears around the centerline of the specimen. Upon further loading, more cracks continue to appear radially away from the centerline until the midplane is well populated with cracks, save for a boundary layer around the free lateral boundary. For instance, for vulcanizate \texttt{D}, the unfilled rubber for which most results were reported by \cite{GL59}, the post-mortem image of the midplane for a very thin specimen with diameter-to-thickness ratio $D/H=33$ shows the presence of about 100 cracks.

    As the diameter-to-thickness ratio $D/H$ decreases, the post-mortem images of the midplanes show that fewer cracks are nucleated. For vulcanizate \texttt{D}, a thick specimen with diameter-to-thickness ratio $D/H=3.5$ shows only one nucleated crack, while the next thicker specimen with diameter-to-thickness ratio $D/H=2$ shows no evidence of cavitation.

    Since no results for the transparent vulcanizate \texttt{G} were reported for thicker specimens, the location and the sequence of nucleation events of the various cracks and whether they propagate are unknown for those. The elongated shape of the cracks in the post-mortem images of the midplanes do suggest that they are nucleated away from the midplane (this much was mentioned by \cite{GL59} about specimens with $D/H<4$), although still around the centerline, and that they propagate significantly with further loading.}

\item{\emph{The when}. From the experimental observations of \cite{GL59}, only the measurements of the normalized global force $S=4P/(\pi D^2)$ and the normalized global deformation $\lambda=h/H$ provide quantitative information about when cavitation occurs. From these measurements (see Subsection \ref{Sec: Comment} below), they noted a linear relation between the first local maximum in the plot $S$ \emph{vs}. $\lambda$, labeled $S^{\prime}$, and the initial shear moduli $\mu$ of the various elastomers that they tested. To determine precisely when cavitation occurs, however, a theoretical analysis is needed. The elasticity analysis initiated by \cite{GL59} is fully reviewed in Section \ref{Sec: Elasticity Theory} below.}

\end{itemize}

Following in the footsteps of \cite{GL59}, numerous researchers made use of poker-chip experiments to continue studying cavitation in a variety of elastomers. Among these, some incorporated innovations in the poker-chip experiment itself which led to new insights into the phenomenon.

\cite{Lindsey67} carried out poker-chip experiments on a transparent polyurethane elastomer by making use of transparent stiff (PMMA) fixtures, in contrast to the opaque metallic fixtures used by his predecessors. While \cite{Lindsey67} restricted attention to specimens of a sole diameter-to-thickness ratio ($D/H=20$), the transparency of the fixtures allowed him to directly observe, for the first time, the appearance of cracks within the specimens during the loading process from an axial perspective.

\cite{Creton10} also carried out poker-chip-like experiments on transparent polyurethane elastomers (of three different cross-link densities and chain lengths) by making use of transparent stiff (glass) fixtures with the added modifications that the top fixture was of spherical --- as opposed to planar --- shape and that the tests were carried out over a range of different temperatures. The spherical shape of the top fixture served to alleviate the demands of the required strong bonding between the elastomer and the fixture as well as to promote the nucleation of fewer cracks in the specimen.

In an effort to generate more comprehensive direct observations, \cite{Bayraktar08} carried out poker-chip experiments using \emph{in situ} X-ray tomography to map out the initial and evolving geometry of the nucleated cracks during the loading process. They restricted attention to thin specimens with a unique (although unspecified) diameter-to-thickness ratio $D/H$ made of natural rubber and SBR both filled with carbon black. Much like the results of \cite{GL59} for their unfilled transparent vulcanizate \texttt{G}, the X-ray tomography results of \cite{Bayraktar08} show that the sequence of events is such that a few cracks first appear near the midplane and around the centerline of the specimens and, upon further loading, more cracks continue to appear away from the centerline until the midplane is populated with cracks, except, again, for a boundary layer around the free lateral boundary.

In a more recent contribution, \cite{Euchler20} also carried out poker-chip experiments on SBR making use of X-ray tomography, at the higher spatial resolution of 10 $\mu$m, in conjunction with dilatometry measurements. In contrast to restricting attention to specimens with a sole diameter-to-thickness ratio $D/H$, as done by  \cite{Bayraktar08}, \cite{Euchler20} considered specimens with four different ratios, $D/H=3.2,10,20,$ and $40$, although X-ray tomography images were generated  only for those with ratios $D/H=10$ and $20$. Consistent with the observations of \cite{GL59} for natural rubber, the X-ray tomography results of \cite{Euchler20} for SBR show that fewer cracks are nucleated when the diameter-to-thickness ratio $D/H$ of the specimen decreases. However, in stark contrast to the observations of \cite{GL59}, the X-ray tomography results of \cite{Euchler20} show that the first nucleated cracks are located at a significant distance --- roughly $D/4$ --- away from the centerline of the specimens (see Fig. \ref{Fig14}(a) below and Fig. 7 in their paper).

Most recently, as part of the experimental program initiated by \cite{Poulain17}, \cite{GuoRavi23} carried out a large number of poker-chip experiments, making use of specimens with diameter-to-thickness ratios in the wide range $D/H\in[2,75]$, on lightly cross-linked silicone elastomers (PDMS Sylgard 184) with one key innovation. Taking advantage of the transparency of the silicone elastomers and the transparency of the fixtures, made of PMMA, the tests were conducted under a high-speed optical microscope, with a temporal resolution of $0.067$ s and a spatial resolution of $1$ $\mu$m, which allowed them to gain new direct insight at an unprecedented spatiotemporal resolution.  Similar to the earlier observations on natural rubber \citep{GL59} and SBR \citep{Euchler20}, the results of \cite{GuoRavi23} for silicone elastomers show that fewer cracks are nucleated when the diameter-to-thickness ratio $D/H$ of the specimen decreases. Interestingly, for specimens with a sufficiently small diameter-to-thickness ratio ($D/H<5$), cracks first nucleate on the lateral free boundary, near one of the fixtures, instead of in the interior of the specimen. In contrast to the earlier observations on natural rubber \citep{GL59}, but consistent with those on SBR \citep{Euchler20}, the results of \cite{GuoRavi23} also show that the first nucleated cracks are \emph{always} located at a significant distance --- between $D/12$ and $D/3$ for internal cracks --- away from the centerline of the specimens, irrespective of their diameter-to-thickness ratio $D/H$. What is more, the first nucleated cracks are also consistently located away --- at a distance between $H/10$ and $H/3$ for internal cracks --- from the midplane (see Fig. \ref{Fig14}(b) below and Figs. 18 and 19 in their paper).

\begin{remark}\label{R: location}
\emph{As we elaborate in Section \ref{Sec: Theory vs Experiments} below, the observations reported by \cite{Euchler20} and \cite{GuoRavi23} on the location of the first nucleated cracks being significantly away from the centerline of the specimens is one of the telltale signs that cavitation cannot possibly be a purely elastic phenomenon.}
\end{remark}

\subsection{Cavitation around filler particles}

\begin{figure}[h!]
\centering
\centering\includegraphics[width=0.65\linewidth]{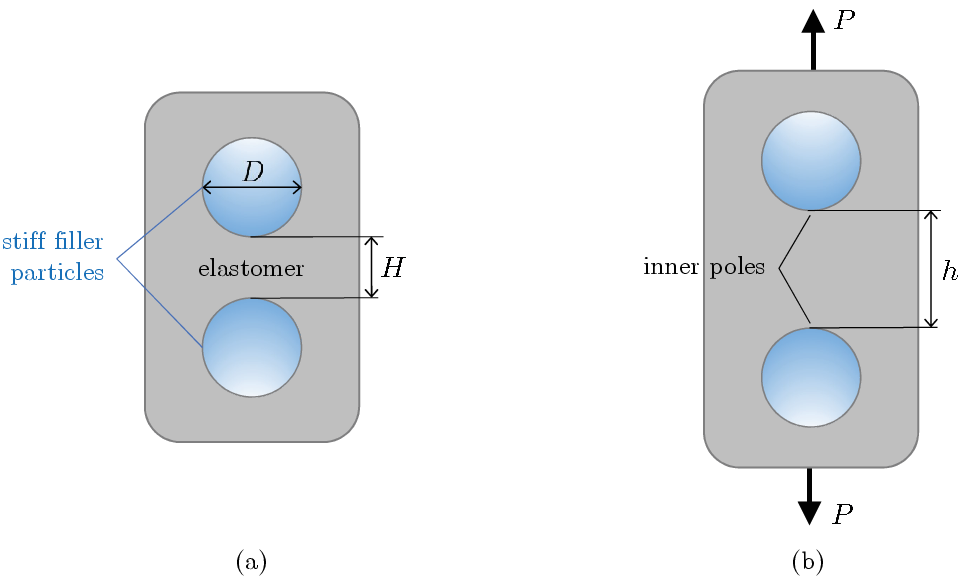}
\caption{\small Schematic of a common experimental setup based on two adjacent filler particles of spherical shape in (a) the initial configuration and in (b) a deformed configuration at a distance $h$ between the particles and corresponding tensile force $P$. Relative to their diameter $D$, the initial distance $H$ between the particles is typically in the range $H/D\in[0.01,0.5]$. For reference in the text, the inner poles of the particles are marked in part (b).}\label{Fig2}
\end{figure}

Another popular type of experiment used to study cavitation is one where stiff filler particles are embedded in the elastomer and the two are pulled apart either directly or indirectly \citep{Oberth65,GentPark84,ChoGent87,ChoGent88,Ilseng17,Poulain17,Poulain18}. Figure \ref{Fig2} provides a schematic of one of the typical experimental setups, where two filler particles of spherical shape are embedded close to one another. This experiment is similar to the poker-chip experiment (\emph{cf}. Fig. \ref{Fig1}) in that a narrow region of the elastomer happens to be placed between two stiff domains, in this case of spherical as opposed to planar shape.

The first cavitation experiments involving filler particles appear to be those of \cite{Oberth65}. These authors embedded a row of spherical particles made of steel (as well as of glass and polystyrene) of about $D=5$ mm in diameter in transparent polyurethane and polybutadiene elastomers, cast the mixture into dog-bone specimens with a cross section of about $10$ mm $\times$ $10$ mm, and then subjected these to uniaxial tension. They reported two types of observations: the global force-deformation response of the specimens and, most importantly,  images at select instances during the loading process of regions around the particles that show the sudden appearance of optically visible cracks near the interfaces of the elastomers with the particles.

About two decades later, \cite{GentPark84} repeated the same type of experiment. In particular, they considered specimens made of various transparent elastomers (natural rubber, polybutadiene, and silicone) filled with either one or two glass spherical particles, among others, of diameters $D=610$ $\mu$m and $1.25$ mm. For the case of one particle, in agreement with the earlier observations of \cite{Oberth65}, their results show the appearance of cracks near the poles of the particle. The experiments with two filler particles turned out to be more interesting. For those, \cite{GentPark84} reported results for two specimens made of the same silicone elastomer containing two particles of the same diameter $D=1.25$ mm separated by a small initial distance, $H/D=0.116$ in one specimen and $H/D=0.152$ in the other. The results pertain to optical images at four select instances during the loading process. Rather intriguingly, these images show first the appearance of two small ($\sim 0.03$ mm) internal cracks near the inner poles of both particles followed by their apparent disappearance together with the appearance of a large crack ($>0.65$ mm) between the two particles.

About five years ago, \cite{Poulain17,Poulain18} revisited the two-particle experiment carried out by \cite{GentPark84} with the main objective of understanding the intriguing ``disappearance'' of the small cracks near the inner poles of the particles and the appearance of the large crack between the two particles. Generalizing the original experiments of \cite{GentPark84}, the experiments were carried out in specimens made of transparent silicone elastomers (PDMS Sylgard 184) of six different cross-link densities containing two glass particles of the same diameter $D=3.170\pm0.009$ mm separated by a plurality of small initial distances in the range $H/D\in[0.012,0.400]$. In addition to considering a wider range of material behaviors and initial particle-to-particle  distances, the experiments of \cite{Poulain17,Poulain18} made use of another key innovation, to wit, the tests were conducted under a high-speed optical microscope with a spatiotemporal resolution of $1$ $\mu$m and $0.067$ s. This resolution was high enough to provide access to the location and the sequence of nucleation events of all the various cracks.

In a nutshell, the results of \cite{Poulain17,Poulain18} show that, irrespective of the initial particle-to-particle distance $H/D$, a small crack of about 10 $\mu$m in size first suddenly appears near one of the inner poles of the particles. For very small initial particle-to-particle distances $H/D<0.1$, upon further loading, more cracks continue to nucleate near the inner poles of both particles, while the previously nucleated cracks exhibit limited propagation. At some point upon further loading, one of the cracks suddenly starts to propagate relatively fast while the remaining cracks shrink in size and disappear from view\footnote{Through multiple cyclic loading, unloading, and reloadings of the specimens, \cite{Poulain18} established that the disappearance of cracks from view corresponds in fact to their healing.}. The propagation of the growing crack eventually leads to the complete rupture of the specimen. For specimens with a larger particle-to-particle distance $H/D>0.1$, upon further loading, the sole small crack nucleated near one of the inner poles of the particles propagates somewhat before another small crack appears near the inner pole of the opposite particle. At that point, one of the two cracks (not necessarily the one that nucleated first) starts to propagate relatively fast while the other crack shrinks in size and disappears from view. In this case too the propagation of the growing crack eventually leads to the complete rupture of the specimen. This latter sequence of events was the one that \cite{GentPark84} partially captured in their experiments. Indeed, the higher spatiotemporal resolution used by \cite{Poulain17,Poulain18} revealed that the large crack that \cite{GentPark84} observed between the two particles is in fact the result of the propagation of one of the cracks that nucleated near the poles, and \emph{not} a separate nucleation event.

In addition to studies that make use of spherical filler particles, there have been studies that have considered filler particles of cylindrical and other anisotropic shapes \citep{ChoGent87,ChoGent88,Ilseng17}. The outcomes are fundamentally the same as for spherical particles: cavitation occurs near the interface of the fillers with the elastomer.

\subsection{Cavitation within inclusions of elastomer embedded in a hard brittle polymer}

\begin{figure}[t!]
\centering
\centering\includegraphics[width=0.7\linewidth]{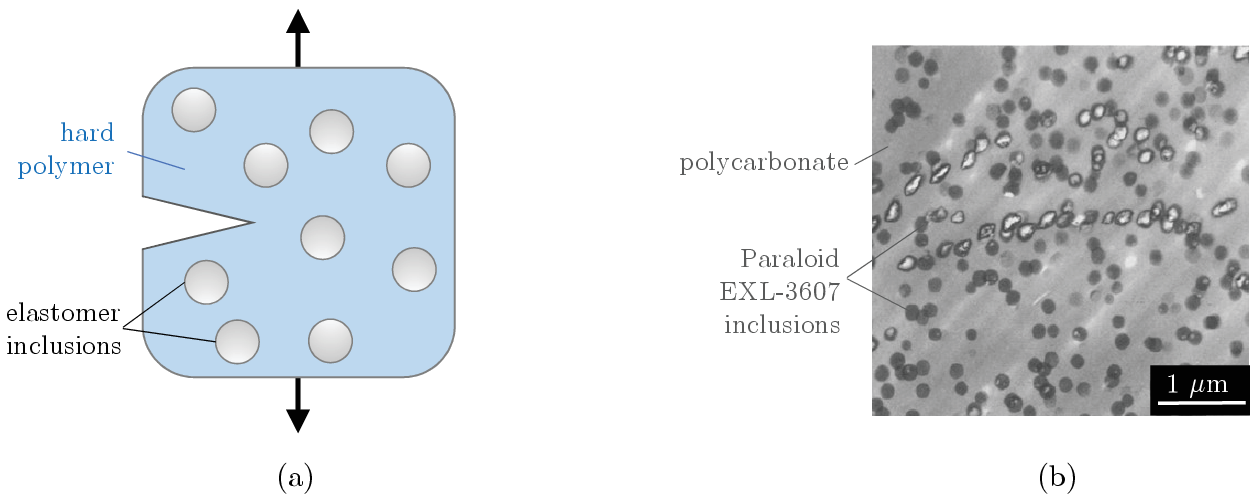}
\caption{\small (a) Schematic of a hard polymer filled with inclusions made of elastomer around a crack front. (b) Representative TEM in a deformed configuration of a polycarbonate polymer filled with 5\% volume fraction of Paraloid EXL-3607 core-shell inclusions of 0.2 $\mu$m in initial diameter illustrating cavitation (shown as a white core) in some of the inclusions \citep{Cheng95}.}\label{Fig3}
\end{figure}

Since the 1970s, a common approach to enhance the macroscopic toughness of hard brittle polymers has been to fill them with a small volume fraction of inclusions made of elastomer \citep{Sultan73,Kinloch83,Pearson91,Cheng95,Pearson09,Kinloch13,Wang19}. This is because ahead of a growing crack in these material systems, where the stress field is triaxial and all-tensile, the inclusions can undergo cavitation, which provides a toughening mechanism; see Fig. \ref{Fig3}(a). The fact that cavitation occurs within the inclusions has been long established from direct microscopy imaging. By way of an example, Fig. \ref{Fig3}(b) provides a representative transmission electron micrograph (TEM) illustrating the cavitation of Paraloid EXL-3607 core-shell inclusions embedded in a polycarbonate polymer \citep{Cheng95}.

With respect to the experimental observations of cavitation summarized in the two preceding subsections, the vast majority of existing experimental observations of cavitation within inclusions of elastomer embedded in hard brittle polymers is fundamentally different on two counts.

First, traditionally\footnote{See, for instance, the review article of \cite{Wang19} for new trends in the design of toughening elastomer inclusions.} the inclusions utilized to toughen hard brittle polymers are made either of an elastomer that is \emph{not} cross linked or of an elastomer that is also \emph{not} cross linked but that is surrounded by a thin layer of a glassy shell \citep{Pearson09}; the image shown in Fig. \ref{Fig3}(b) corresponds to an example of the latter. That is, the observations pertain to cavitation of elastomers in a \emph{liquid} state, as opposed to those summarized in the two preceding subsections that pertain to cross-linked elastomers.

Second, the diameters of the inclusions that are most commonly \citep{Pearson09} utilized to toughen hard brittle polymers are in the range $[0.1,5]$ $\mu$m. Accordingly, the observations pertain to the sudden appearance of internal cracks that are often \emph{submicron} in size, as opposed to those summarized in the two preceding subsections that pertain to internal cracks of size $\geq 1\mu$m.

\section{The elasticity view}\label{Sec: Elasticity Theory}

\subsection{The first result}\label{Sec: The first result}

Out of all the experimental observations that they generated from their poker-chip experiments, \cite{GL59} chose to focus on explaining the first instance at which an optically visible crack suddenly appears in the specimens\footnote{Here, it is important to emphasize that the poker-chip experiments of \cite{GL59} provided a wealth of observations beyond the fact that optically visible cracks suddenly appear in the specimens. Again, they provided observations of the location and sequence of nucleation events, they also showed that the number of nucleated cracks decreases when the diameter-to-thickness ratio $D/H$ decreases, that for smaller $D/H$ the shape of the cracks can be very elongated suggesting that they undergo significant propagation, and that the global $S$ \emph{vs}. $\lambda$ response of the specimens exhibits an initial stiffening region, followed by a plateau, followed by a secondary stiffening region before complete rupture is reached. Strikingly, these additional observations have been ignored by the majority of studies of cavitation in the literature.}. They hypothesized that such an event is due to the \emph{elastic growth} of the inherent defects in the elastomers when the material around these defects experiences a sufficiently large tensile hydrostatic stress.

\begin{figure}[t!]
\centering
\centering\includegraphics[width=0.7\linewidth]{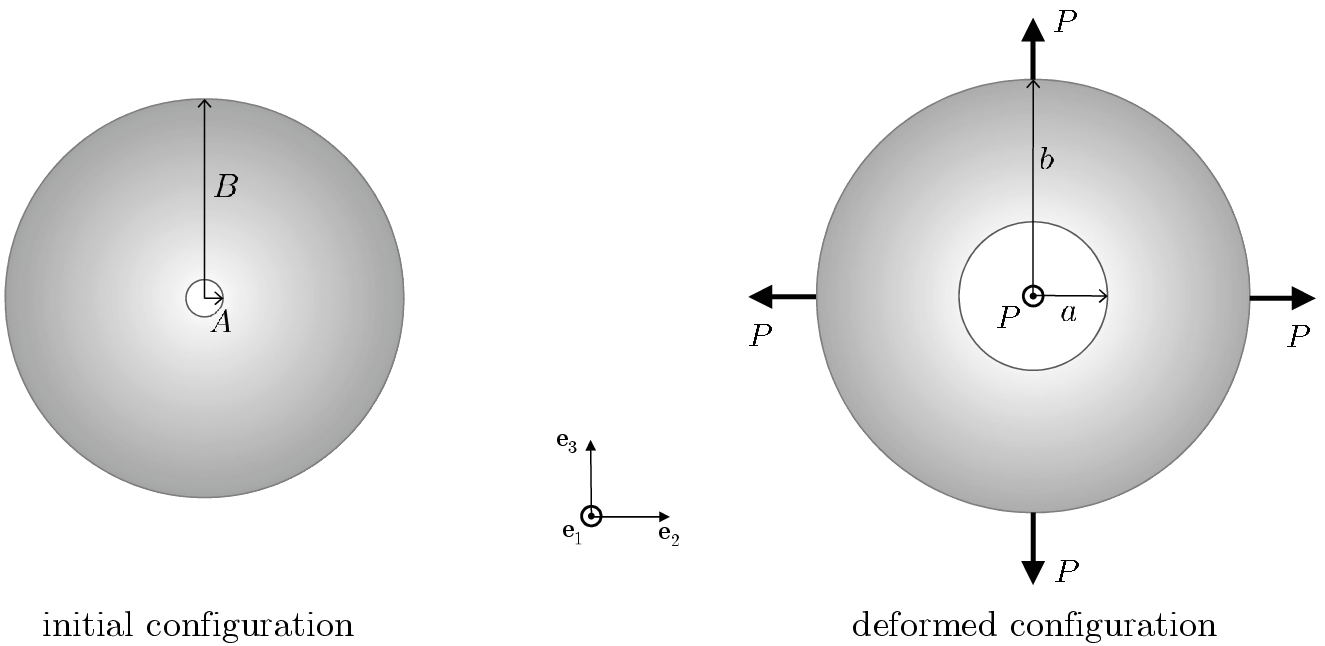}
\caption{\small Schematic of the boundary-value problem --- of a vacuous spherical cavity of initial radius $A$ embedded at the center of a ball of initial radius $B\gg A$ made of an incompressible isotropic nonlinear elastic solid subjected to a tensile Cauchy pressure $P$ on its outer boundary --- considered by \cite{GL59} in their elastic analysis of cavitation.}\label{Fig4}
\end{figure}
Acting on their hypothesis, \cite{GL59} considered the elementary elastostatics boundary-value problem schematically depicted in Fig. \ref{Fig4}, that of a vacuous spherical cavity of initial radius $A$, embedded at the center of a ball of initial radius $B\gg A$ made of an incompressible isotropic nonlinear elastic solid, with stored-energy function $W=W(\lambda_1,\lambda_2,\lambda_3)$ in terms of the principal stretches $\lambda_1$, $\lambda_2$, $\lambda_3$, subjected to a tensile Cauchy pressure $P$ on its outer boundary. By restricting attention to radially symmetric deformations and specializing the general relation \citep{Green-Zerna}
\begin{align}
P&=\displaystyle\int_{\left(1+\frac{a^3-A^3}{B^3}\right)^{1/3}}^{\frac{a}{A}}\dfrac{1}{z^3-1}\dfrac{{\rm d}W}{{\rm d}z}\left(z^{-2},z,z\right)\,{\rm d}z\nonumber\\
&=\displaystyle\int_{\frac{b}{B}}^{\left(1+\frac{b^3-B^3}{A^3}\right)^{1/3}}\dfrac{1}{z^3-1}\dfrac{{\rm d}W}{{\rm d}z}\left(z^{-2},z,z\right)\,{\rm d}z\label{P-a}
\end{align}
between the applied pressure $P$ and the resulting radius $a$ ($b$) of the cavity (ball) in the deformed configuration to a Neo-Hookean material with stored-energy function
\begin{equation}
W=\left\{\begin{array}{ll}\dfrac{\mu}{2}\left(\lambda_1^2+\lambda_2^2+\lambda_3^2-3\right) & {\rm if}\; \lambda_1 \lambda_2\lambda_3=1 \vspace{0.1cm}\\
+\infty & {\rm else}\end{array}\right.,\label{W-NH}
\end{equation}
\cite{GL59} showed that there is a critical value $P_c$ of the applied pressure $P$ at which a cavity of infinitesimally small radius $A=0+$ grows suddenly to finite size through an instability; recall that the parameter $\mu$ in (\ref{W-NH}) stands for the initial shear modulus of the material. That critical value is given by the now-famous result
\begin{equation}
P_{c}=\displaystyle\int_{1}^{+\infty}\dfrac{2\mu}{z^3-1}\left(z-z^{-5}\right)\,{\rm d}z=\dfrac{5}{2}\mu.\label{Pc-Neo}
\end{equation}
In view of this result, \cite{GL59} \emph{de facto} postulated \citep{Gent90} that an internal crack will suddenly appear at a material point in an elastomer whenever the hydrostatic part $\sigma_m:={\rm tr}\,\bfT\,/3=(t_1+t_2+t_3)/3$ of the Cauchy stress $\bfT$ at that point reaches the critical value (\ref{Pc-Neo}), to wit, in the form conventionally used to define failure envelopes, whenever the criticality condition
\begin{equation}
\dfrac{1}{3}(t_1+t_2+t_3)-P_{c}=0,\quad {\rm with} \quad P_{c}=\dfrac{5}{2}\mu,\label{CC-GL}
\end{equation}
is first satisfied. With the introduction of the criticality condition (\ref{CC-GL}), the elasticity view of the phenomenon of cavitation in elastomers was born.

\begin{remark}
\emph{The criterion (\ref{CC-GL}) cannot be correct in general. For instance, when specialized to uniaxial tension with $t_2=t_3=0$ and $t_1>0$, the criterion (\ref{CC-GL}) predicts that cavitation occurs at $t_1=15\mu/2$, when in fact there is no experimental evidence that cavitation occurs under uniaxial tension. This simple observation has often been overlooked in the literature.}
\end{remark}

\subsection{The general result}

The derivation of the criticality condition (\ref{CC-GL}) makes use of three significant simplifying idealizations:
\begin{itemize}

\item{The behavior of the elastomer is Neo-Hookean.}

\item{The loading is purely hydrostatic.}

\item{There is only one defect, which is modeled as a vacuous cavity of initially spherical shape and infinitesimally small size.}

\end{itemize}
In the sequel, we provide a summary of the results that go beyond these idealizations, one at a time, and lead to the general elasticity view of the phenomenon of cavitation in elastomers.

Much like in the preceding section on experiments, here too we strive to provide some historical perspective by presenting the results in chronological order, as they
have been reported in the literature since the 1950s until the 2010s, when the theory was substantially completed.

\subsubsection{Beyond Neo-Hookean behavior}\label{Sec:Beyond NH}

Within the restricted setting of purely hydrostatic loading and a single defect of spherical shape and infinitesimally small size, the generalization of (\ref{CC-GL}) to account for an arbitrary (suitably well-behaved) incompressible isotropic stored-energy function $W(\lambda_1,\lambda_2,\lambda_3)$ is straightforward. This is because one can still make use of the general relation (\ref{P-a}) --- since it applies to arbitrary  stored-energy functions $W(\lambda_1,\lambda_2,\lambda_3)$ --- to deduce that the critical value $P_c$ of the applied pressure $P$ at which a cavity of infinitesimally small radius $A=0+$ grows to finite size is given by the formula
\begin{equation}
P_{c}=\displaystyle\int_{1}^{+\infty}\dfrac{1}{z^3-1}\dfrac{{\rm d}W}{{\rm d}z}\left(z^{-2},z,z\right)\,{\rm d}z.\label{Pc}
\end{equation}
\begin{remark}
\emph{The upper limit of integration in the formula (\ref{Pc}) being $+\infty$ indicates that the critical value $P_c$ depends fundamentally on the specifics of how the elastomer behaves elastically at infinitely large stretches; in other words, different assumed behaviors of $W$ at infinitely large stretches can lead to wildly different values for $P_{c}$. This is clearly incongruent with the fact that no material can sustain infinitely large elastic stretches and hence a hint that cavitation is not an elastic phenomenon.}
\end{remark}

The result (\ref{Pc}) was established rigorously by \cite{Ball82}. In his celebrated contribution, \cite{Ball82} also established two other results with profound implications. The first one is that the elasticity view of cavitation as the sudden growth of a defect is equivalent to the elasticity view of cavitation as a \emph{non-smooth} bifurcation in an initially ``perfect'' material without defects; see also the work of \cite{Sivaloganathan86}. The second one is that the improper integral in (\ref{Pc}) is \emph{unbounded} for stored-energy functions with growth conditions such that
\begin{equation}
\dfrac{W(\lambda_1,\lambda_2,\lambda_3)}{(\lambda_1^2+\lambda_2^2+\lambda_3^2)^{3/2}}\nearrow +\infty \quad {\rm as}\quad \lambda_1^2+\lambda_2^2+\lambda_3^2\nearrow+\infty;\label{Growth}
\end{equation}
see also the work of \cite{MS95}. In other words, the critical pressure (\ref{Pc}) does not converge to a finite value for an elastomer whose elasticity is characterized by a stored-energy function $W$ that increases faster than the norm $(\lambda_1^2+\lambda_2^2+\lambda_3^2)^{3/2}$ at large stretches.
\begin{remark}\label{R: p}
\emph{The growth condition (\ref{Growth}) rules out the possibility of cavitation in elastomers with standard non-Gaussian stiffening at large stretches. Accordingly, the use of  common non-Gaussian elasticity models would imply the impossibility of cavitation. These include, for instance, the Mooney-Rivlin model \citep{Mooney40}, the Ogden model with a power larger than 3 in the stretches $\lambda_1$, $\lambda_2$, $\lambda_3$ \citep{Ogden72}, the Yeoh model or any other more general polynomial model with a power equal to or larger than $2$ in the first invariant terms $(I_1-3)$ \citep{Yeoh93}, any model with limiting chain extensibility, such as the Arruda-Boyce and the Gent models \citep{AB93,Gent96}, and the Lopez-Pamies model with a power larger than $3/2$ in the first invariant $I_1$ \citep{LP10}, among many others. We will come back to this pivotal result in Section \ref{Sec: Theory vs Experiments} below, when we confront the elasticity view of cavitation with experimental observations.}
\end{remark}

The generalization of (\ref{CC-GL}) to account for an arbitrary (suitably well-behaved) \emph{compressible} isotropic stored-energy function $W(\lambda_1,\lambda_2,\lambda_3)$ requires having to deal directly with the equations of elastostatics --- specialized to radially symmetric deformations --- to determine the critical value $P_c$ of the applied pressure $P$ at which a cavity of infinitesimally small radius $A=0+$ grows to finite size \citep{Ball82,Stuart85}. While this is also conceptually straightforward, it requires in general a fully numerical treatment of the problem to compute $P_c$.

\subsubsection{Beyond hydrostatic loading}

\begin{figure}[H]
\centering
\centering\includegraphics[width=0.7\linewidth]{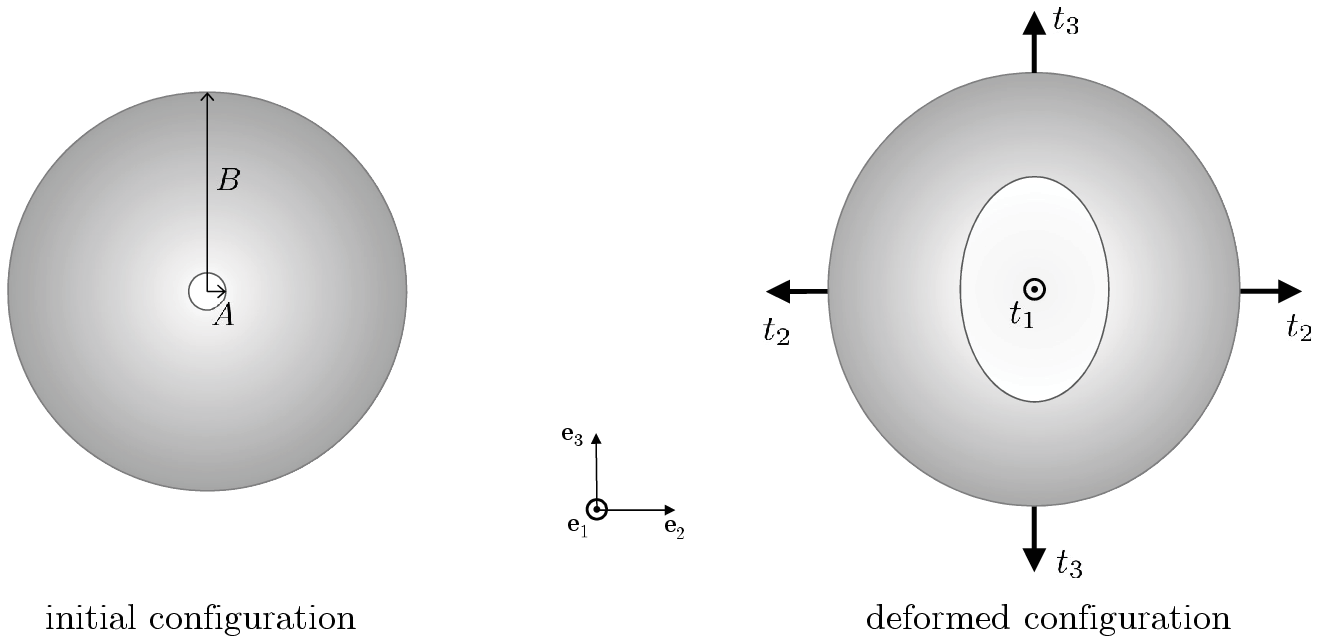}
\caption{\small Schematic of the boundary-value problem of a vacuous spherical cavity of initial radius $A$ embedded at the center of a ball of initial radius $B\gg A$ made of a nonlinear elastic solid subjected to a general affine Cauchy stress $\bfT={\rm diag}(t_1,t_2,t_3)$ on its outer boundary.}\label{Fig5}
\end{figure}

Still within the restricted setting of a single defect of spherical shape and infinitesimally small size, the generalization of (\ref{CC-GL}) to account for arbitrary loading conditions is conceptually clear. As depicted schematically by Fig. \ref{Fig5}, it suffices to consider the same boundary-value problem considered above with the alteration that the ball is now subjected to an affine Cauchy stress of the general triaxial form $\bfT={\rm diag}(t_1,t_2,t_3)$, instead of just $\bfT={\rm diag}(P,P,P)$, on its outer boundary. In general, this problem does not admit analytical solutions and hence must be solved numerically.

As a first step to gain quantitative insight, \cite{Abeyaratne92} made use of a clever kinematically admissible trial field to determine an analytical approximation for the set of critical values of the applied principal Cauchy stresses $(t_1,t_2,t_3)$ at which a cavity of infinitesimally small radius $A=0+$ grows to finite size in a Neo-Hookean elastomer. That set of critical values defines a surface in the space of principal Cauchy stresses $(t_1,t_2,t_3)$ given by
\begin{equation}
(4 t_1-t_2-t_3)(4 t_2-t_3-t_1)(4 t_3-t_1-t_2)-125 \mu^3=0\label{HA-Criterion}
\end{equation}
with $4 t_1-t_2-t_3>0$, $4 t_2-t_3-t_1>0$, and $4 t_3-t_1-t_2>0$. By construction, this surface is an upper bound, that is, the sudden elastic growth of a vacuous spherical  cavity of infinitesimal size in a Neo-Hookean elastomer will occur strictly before any of the stress states defined by (\ref{HA-Criterion}) is reached, save for the case of hydrostatic stress states with $t_1=t_2=t_3=P$ for which (\ref{HA-Criterion}) reduces to (\ref{CC-GL}) and hence it is an exact result.

A year after \cite{Abeyaratne92} reported the upper bound (\ref{HA-Criterion}), \cite{Gent93} worked out corresponding numerical approximations by means of the finite element (FE) method. Their results pertain to a nearly incompressible Neo-Hookean material, one with initial bulk-to-shear-modulus ratio $\kappa/\mu=500$, and axisymmetric loading with $\bfT={\rm diag}(t_1,t_2,t_3=t_2)$. Because \cite{Gent93} made use of a standard deformation-based FE method, their results suffer from volumetric locking and hence are \emph{not} accurate converged solutions.

\begin{figure}[t!]
\centering
\centering\includegraphics[width=0.55\linewidth]{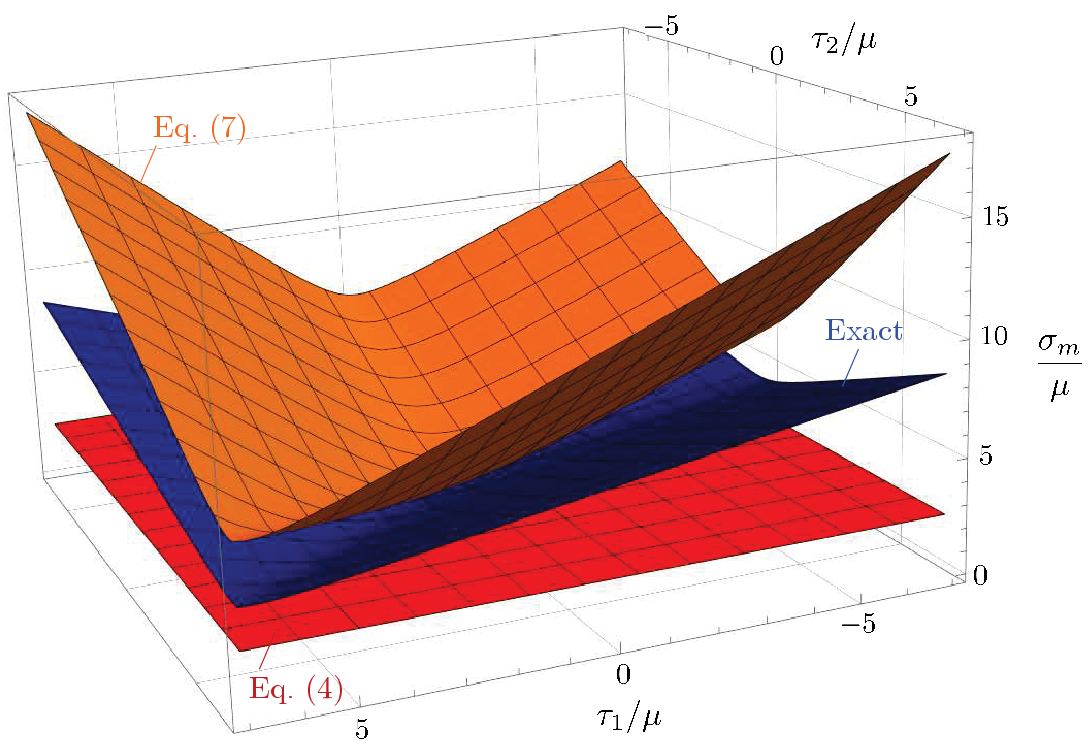}
\caption{\small Plot of the critical stresses in the normalized shear-hydrostatic space $(\tau_1/\mu,\tau_2/\mu,\sigma_{m}/\mu)$ at which a cavity of infinitesimally small radius $A=0+$ grows to finite size in a Neo-Hookean elastomer \citep{LP11b}. For direct comparison, the figure includes the classical criterion (\ref{CC-GL}) of \cite{GL59} and the upper bound (\ref{HA-Criterion}) of \cite{Abeyaratne92}.}\label{Fig6}
\end{figure}

Converged FE solutions for the basic case of a Neo-Hookean elastomer under general triaxial loadings $\bfT={\rm diag}(t_1,t_2,t_3)$ were worked out almost two decades later, in 2011, by \cite{LP11b}. These authors made use of a hybrid FE formulation, which allowed them to deal with the incompressibility constraint (\ref{W-NH})$_2$ of the material in an exact manner. Their results for the set of critical values of the applied principal Cauchy stresses $(t_1,t_2,t_3)$ at which a cavity of infinitesimally small radius $A=0+$ grows to finite size are reproduced in Fig. \ref{Fig6}. To help visualize the effect of the stress triaxiality, they are plotted in the normalized space $(\tau_1/\mu,\tau_2/\mu,\sigma_{m}/\mu)$, where
\begin{equation*}
\quad \tau_1:=t_2-t_1,\quad \tau_2:=t_3-t_1,\quad \sigma_{m}=\frac{1}{3}(t_1+t_2+t_3)
\end{equation*}
stand for the shear and hydrostatic parts of the stress. For direct comparison, the figure includes the classical criterion (\ref{CC-GL}) of \cite{GL59} and the upper bound (\ref{HA-Criterion}) of \cite{Abeyaratne92}.

Two key observations are immediate from Fig. \ref{Fig6}. First, the critical stresses at which the cavity grows to finite size are all tensile, namely, they are such that $t_1=\sigma_m-(\tau_1+\tau_2)/3>0$, $t_2=\sigma_m+(2\tau_1-\tau_2)/3>0$, and $t_3=\sigma_m-(\tau_1-2\tau_2)/3>0$. Second, the critical stresses are strongly dependent on the value of the shear stresses $\tau_1$ and $\tau_2$ and \emph{not} just solely on the hydrostatic part $\sigma_{m}$ of the stress. In particular, the larger the deviation from the purely hydrostatic stress (i.e., the larger $|\tau_1|$ and $|\tau_2|$), the larger the critical value of  $\sigma_{m}$ at which the cavity grows. A direct implication of this behavior is that the classical criterion (\ref{CC-GL}) of \cite{GL59} bounds from below the exact solution. Figure \ref{Fig6} also serves to illustrate that the upper bound (\ref{HA-Criterion}) deviates significantly from the exact solution as $|\tau_1|$ and $|\tau_2|$ increase.

In a companion work aimed at gaining insight into the effect of stress triaxiality in elastomers with general types of incompressible isotropic elastic behaviors, beyond Neo-Hookean, \cite{NLP12} worked out results analogous to those presented by \cite{LP11b} for elastomers with various types of stored-energy functions $W$, including the one-term Odgen
\begin{equation}
W=\left\{\begin{array}{ll}\dfrac{2\mu}{\beta^2}\left(\lambda_1^{\beta}+\lambda_2^{\beta}+\lambda_3^{\beta}-3\right) & {\rm if}\; \lambda_1 \lambda_2\lambda_3=1 \vspace{0.1cm}\\
+\infty & {\rm else}\end{array}\right.\label{W-O}
\end{equation}
and the one-term Lopez-Pamies
\begin{equation}
W=\left\{\begin{array}{ll}\dfrac{3^{1-\alpha}\mu}{2\alpha}\left((\lambda_1^2+\lambda_2^2+\lambda_3^2)^{\alpha}-3^{\alpha}\right) & {\rm if}\; \lambda_1 \lambda_2\lambda_3=1 \vspace{0.1cm}\\
+\infty & {\rm else}\end{array}\right.\label{W-LP}
\end{equation}
stored-energy functions. As alluded to in Remark \ref{R: p} above, the choices of stored-energy functions (\ref{W-O}) and (\ref{W-LP}) are particularly convenient because their growth conditions at large stretches can be modulated by simply changing the values of the material constants $\beta$ and $\alpha$; note that when $\beta=2$ and $\alpha=1$, both (\ref{W-O}) and (\ref{W-LP}) reduce identically to the Neo-Hookean stored-energy function (\ref{W-NH}). So as to ensure finite values for the critical stresses $(t_1,t_2,t_3)$ at which the cavity grows to finite size, \cite{NLP12} studied the cases $\beta=\{1.5,1\}$ and $\alpha=\{0.8,0.6\}$, all of which correspond to ``sub-Neo-Hookean'' growth conditions with
\begin{equation*}
\dfrac{W(\lambda_1,\lambda_2,\lambda_3)}{(\lambda_1^2+\lambda_2^2+\lambda_3^2)^{3/2}}\searrow 0\quad {\rm as}\quad \lambda_1^2+\lambda_2^2+\lambda_3^2\nearrow+\infty.
\end{equation*}
Much like for the Neo-Hookean case, the results show that the cavity grows to finite size only when all the three principal Cauchy stresses are tensile. They also show that the critical value of the hydrostatic part $\sigma_m$ of the stress at which the cavity grows depends strongly on the stress triaxiality (as characterized by $\tau_1$ and $\tau_2$), as well as on the growth conditions of the elastic behavior of the material (as characterized by $\beta$ and $\alpha$).

Results similar to those of \cite{NLP12} were also obtained by \cite{Negron12} for a class of \emph{compressible} stored-energy functions with ``sub-Neo-Hookean'' growth.

\subsubsection{Beyond a single defect of initially spherical shape}

Next, the generalization of (\ref{CC-GL}) to account for the presence not just of a single defect but of a distribution of defects leads to the casting of the problem as a \emph{homogenization problem} in finite elastostatics \citep{LP11a}.

\begin{figure}[h!]
\centering
\centering\includegraphics[width=0.7\linewidth]{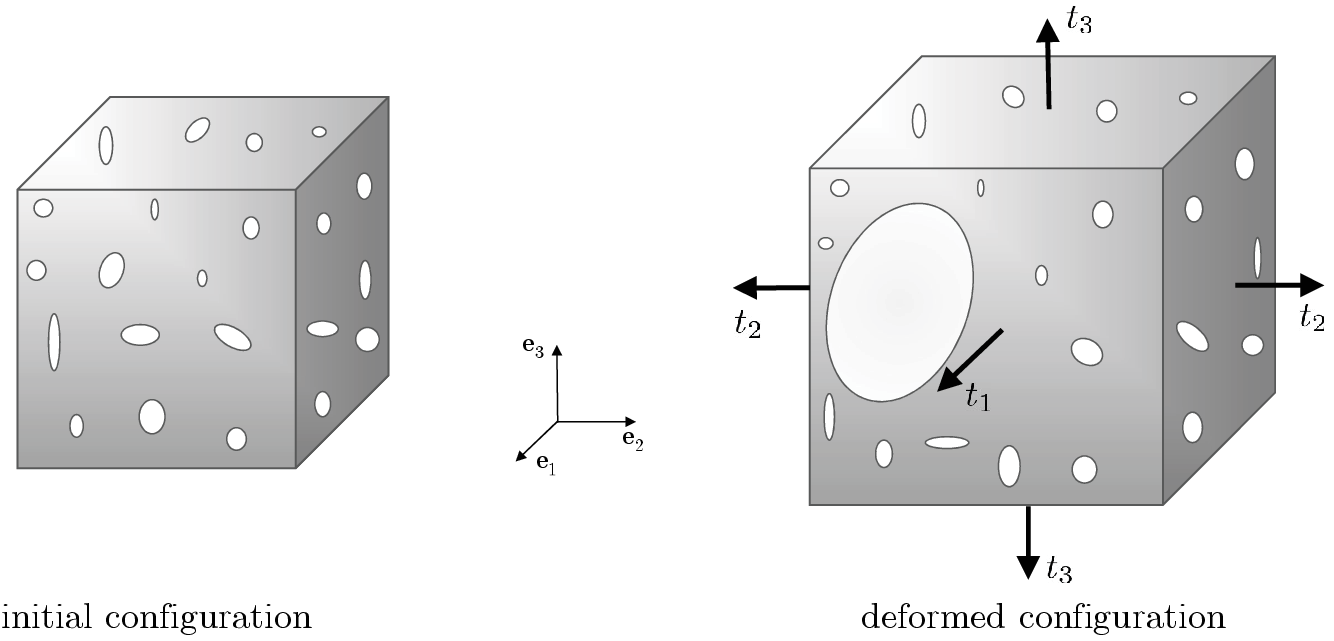}
\caption{\small Schematic of the boundary-value problem of a statistically uniform distribution of vacuous cavities of small size, small initial volume fraction $f_0$, and arbitrary shape embedded in a nonlinear elastic solid subjected to a general affine Cauchy stress $\bfT={\rm diag}(t_1,t_2,t_3)$ on its outer boundary.}\label{Fig7}
\end{figure}
Specifically, as illustrated schematically by Fig. \ref{Fig7}, the problem amounts to considering a nonlinear elastic solid, with stored-energy function $W$, that contains a statistically uniform distribution of vacuous cavities, of any shape of choice, at an infinitesimally small initial volume fraction $f_0=0+$, that is subjected to an affine Cauchy stress of the general triaxial form $\bfT={\rm diag}(t_1,t_2,t_3)$ on its outer boundary. Within this general formulation, cavitation corresponds to the event when at least one of the cavities grows to finite size. That growth is most distinctly manifested by the sudden increase of the current volume fraction $f$ of cavities in the deformed configuration to a finite value \citep{LP09,LP11a}.

Making use of a combination of iterated homogenization techniques \citep{Norris85,LP10a,Tartar85,Francfort86}, \cite{LP11b} worked out an exact homogenization solution for a fairly general class of distributions of cavities in an elastomer characterized by any --- possibly compressible and anisotropic --- suitably well-behaved stored-energy function $W$ of choice. The solution is given in terms of a certain Hamilton-Jacobi equation, which, in general, needs to be solved numerically \citep{LGLP19}.

For the basic case when the cavities are distributed isotropically and the elastomer is Neo-Hookean, the solution of the resulting Hamilton-Jacobi equation can be accurately approximated by a closed-form expression. Based on that approximation, \cite{LP11b} established that infinitesimally small defects that are isotropically distributed in a Neo-Hookean elastomer can first grow to finite size whenever the principal Cauchy stresses $(t_1,t_2,t_3)$ reach the criticality condition
\begin{equation}
8 t_1 t_2 t_3-12\mu(t_1 t_2+t_2 t_3+t_3 t_1)+18\mu^2(t_1+t_2+t_3)-35\mu^3=0\label{IH-Neo}
\end{equation}
with $t_1,t_2,t_3>3\mu/2$.

By means of a direct comparison, \cite{LP11b} determined that the general condition (\ref{IH-Neo}) is both qualitatively and quantitatively similar to the FE results reproduced in Fig. \ref{Fig6} for the critical stresses at which a single vacuous spherical cavity of infinitesimal size in a Neo-Hookean elastomer grows to finite size. What is more, a simple calculation shows that the general condition (\ref{IH-Neo}) reduces identically to the classical condition (\ref{CC-GL}) for the growth of a single vacuous spherical cavity of infinitesimal size in a Neo-Hookean elastomer under purely hydrostatic loading when $t_1=t_2=t_3=P$.

From a fundamental point of view, the agreement between the result (\ref{IH-Neo}) and the result in Fig. \ref{Fig6} for a single spherical defect reveals that the shape of defects and their interaction with one another --- so long as they are infinitesimally small in size --- have a marginal effect on the critical stresses at which they grow to finite size, at least in Neo-Hookean elastomers\footnote{Several mathematical and numerical results \citep{Siva06,Henao09,Henao11,Li11} have suggested that the effects of the shape of infinitesimally small defects and their interaction with one another remain marginal on the critical stresses at which they grow to finite size in elastomers with other types of elastic behaviors, not just Neo-Hookean.}.

From a practical point of view, the agreement also indicates that the criticality condition (\ref{IH-Neo}) can be viewed as the universal criterion for the onset of elastic cavitation in Neo-Hookean elastomers, irrespective of the specific shape and spatial distribution of underlying defects, so long, again, as they are infinitesimally small in size.

\subsubsection{Beyond defects of infinitesimally small size}\label{Sec:Beyond Infinitesimal}

Finally, the idealization that the underlying defects are of an infinitesimally small size --- which is a central premise to all of the results summarized in the preceding subsections --- can be removed at the expense of modeling the defects explicitly.

\begin{figure}[b!]
\centering
\centering\includegraphics[width=0.95\linewidth]{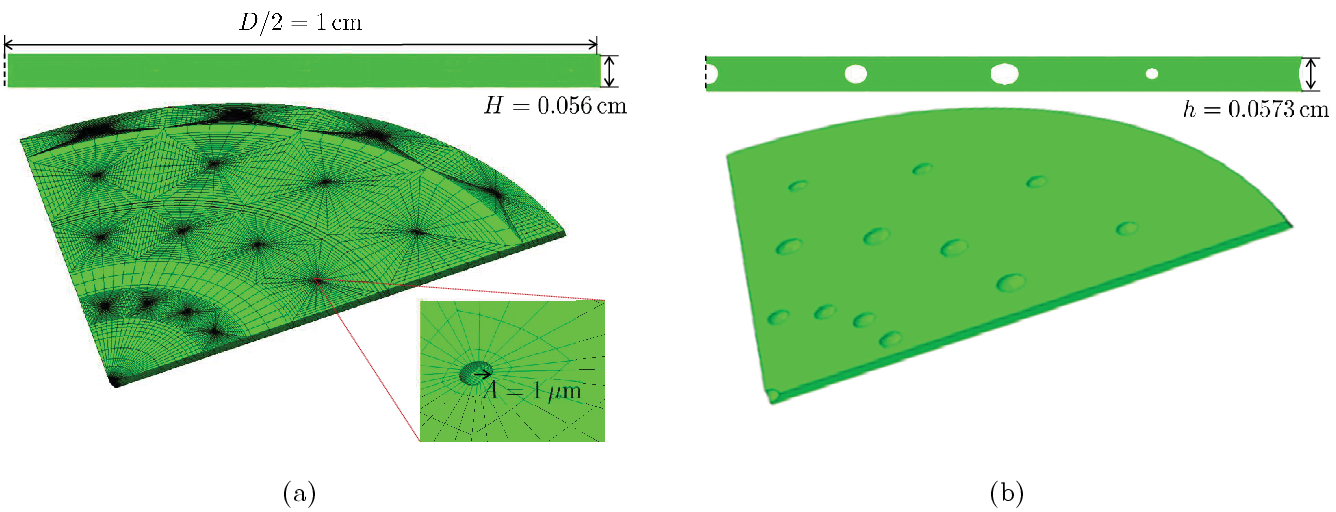}
\caption{\small (a) FE discretization used in the elastic simulation by \cite{LRLP15} of the poker-chip experiment of \cite{GL59} for the specimen with initial thickness $H = 0.056$ cm. The discretization contains a total of 195 defects, which are modeled as vacuous spherical cavities of radius $A=1$ $\mu$m. (b) The deformed configuration of the specimen, modeled as a Neo-Hookean solid, at an applied separation $h=0.0573$ cm between the fixtures illustrating the large growth only of the cavities located in the midplane of the specimen away from its lateral free boundary.}\label{Fig8}
\end{figure}

In 2011 and 2012, a series of 2D works \citep{Henao11,Li11,Li12} studied numerically the response of circular domains, made of a compressible isotropic nonlinear elastic solid, containing two vacuous cavities of various small sizes placed at different locations, that were subjected to a radial stretch $\lambda$ on their outer boundary. The simulations from these works show that cavities that are no larger than $50$ times smaller than the domain occupied by the solid grow to a large size roughly at the same critical stretch $\lambda_c$ at which an infinitesimally small cavity would grow to finite size, provided that the second cavity is sufficiently far away. In the event that the second cavity is not sufficiently far away, depending on their location and size, it may be that only one of the two cavities ends up growing to a large size \citep{Li11,Li12}. That growth also happens roughly at the same critical stretch $\lambda_c$ at which an infinitesimally small cavities would grow to finite size. The recent work of \cite{Henao19} has provided mathematical insight into these 2D numerical results, in particular, into what ``sufficiently far away'' means for two interacting cavities.

In their critical elastic analysis in 2015 of the classical poker-chip experiments of \cite{GL59} and the classical two-particle experiments of \cite{GentPark84}, \cite{LRLP15} modeled explicitly the presence of a distribution of initial defects of small but finite size throughout the specimens. Specifically, they modeled the defects as  vacuous spherical cavities of different radii in the range $A\in[0.2,1]$ $\mu$m. The simulations, carried out for an elastomer with Neo-Hookean behavior, reveal two key results.

First, irrespective of their size, consistent with the 2D results \citep{Li11,Li12} mentioned above, the cavities \emph{start} to rapidly grow roughly when the state of stress around their location reaches the criticality condition (\ref{IH-Neo}), which, once more, was derived under the premise that the underlying defects are infinitesimally small. In other words, the very \emph{onset} of elastic cavitation is rather insensitive to the size of the defects, so long as such a size is small relative to the size of the specimen.

Second, also consistent with the 2D results \citep{Li11,Li12} mentioned above, the simulations show that whether a given defect ends up growing to a ``large'' size in the sense that it is comparable to the size of the specimen depends on its interaction with other defects and the boundaries of the specimen. For instance, as reproduced in Fig. \ref{Fig8} from the simulation by \cite{LRLP15} of the very thin poker-chip experiment with initial thickness $H=0.056$ cm, the cavities that end up growing to a large size are only those that are located in the midplane away from the lateral free boundary. Precisely, the first cavity to grow is the one located at the center of the specimen; more on this in Subsection \ref{Sec: Location} below. Upon further loading, adjacent midplane cavities successively grow in a radial cascading sequence. This observation confirms the expectation that, in a general boundary-value problem, the criticality condition (\ref{IH-Neo}) --- or the corresponding condition for elastomers with non-Neo-Hookean elasticity --- is a necessary but \emph{not} sufficient condition for small defects to grow elastically to large sizes. In order to conclusively determine whether a defect ends up growing to finite size in a given boundary-value problem, one has to solve that boundary-value problem accounting explicitly for the presence of defects.

\section{The elasticity view vs. experimental observations}\label{Sec: Theory vs Experiments}

Having reviewed the experimental observations of cavitation in elastomers and their theoretical description as an elastic phenomenon, we are now in a position to compare the two to pinpoint the reasons why the experimental observations cannot possibly be explained by the elasticity view of cavitation. As announced in the Introduction, there are essentially two reasons:
\begin{enumerate}[label=\roman*.]

\item{The elasticity of elastomers is non-Gaussian.}

\item{The location where cavitation occurs does \emph{not} correlate with the location where defects would grow elastically.}

\end{enumerate}
We discuss them one at a time.

\subsection{The elasticity of elastomers is non-Gaussian}\label{Sec: non-Gaussian}

Because of the finite length of the polymer chains that they are made of, the elasticity of elastomers is necessarily non-Gaussian and hence typically features a distinctly strong stiffening at large stretches \citep{Treloar75}. Precisely, since the early investigations on the mechanics of natural rubber dating back to the 1920s, basic uniaxial and biaxial tension tests have repeatedly shown that the non-Gaussian stiffening of the vast majority of elastomers are characterized by stored-energy functions $W$ with strong growth conditions such that
\begin{equation}
W(\lambda_1,\lambda_2,\lambda_3)>(\lambda_1^2+\lambda_2^2+\lambda_3^2)^{p}\quad {\rm as}\quad \lambda_1^2+\lambda_2^2+\lambda_3^2\nearrow+\infty,\label{Growth-3}
\end{equation}
where the power $p$ is typically in the range
\begin{equation}
p\in[2,10].\label{p growth}
\end{equation}
\begin{remark}\label{R: strength liquids}
\emph{The typical values (\ref{p growth}) of the power $p$ that describes the non-Gaussian stiffening of most elastomers is well above the threshold $p=3/2$ that rules out the possibility of the elastic growth of infinitesimally small defects to finite size; see Remark \ref{R: p} above.}
\end{remark}
%


%
\begin{table}[h!]\centering
\caption{Values of the material constants in the stored-energy function (\ref{W-LP-2-term}) describing the elasticity of three representative elastomers used to study cavitation in the literature.}
\begin{tabular}{ccccc}
\toprule
 & $\mu_1$ (MPa) & $\alpha_1$ & $\mu_2$ (MPa) & $\alpha_2$ \\
\hline
natural rubber \citep{GL59} & $0.583$ & $0.650$ & $0.005$ & $2.725$ \\
silicone \citep{Poulain17} & $0.011$ & $0.518$ & $0.003$ & $2.050$ \\
SBR \citep{Euchler20} & $0.478$ & $0.468$ & $0.005$ & $2.150$ \\
\bottomrule
\end{tabular} \label{Table1}
\end{table}

\begin{figure}[t!]
  \subfigure[]{
   \label{fig:9a}
   \begin{minipage}[]{0.475\linewidth}
   \centering \includegraphics[width=0.9\linewidth]{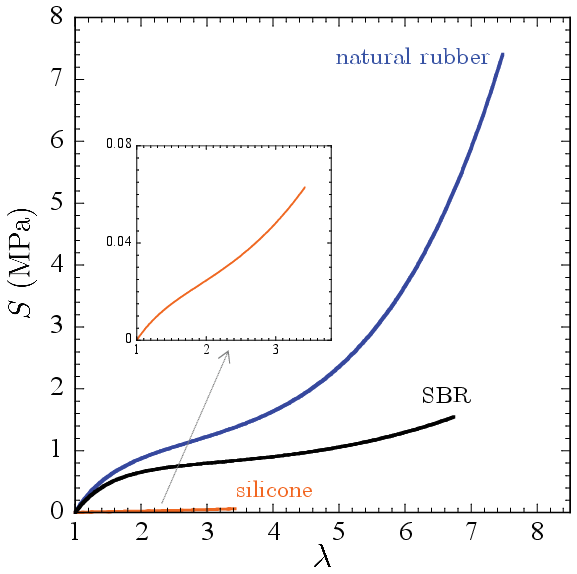}
   \vspace{0.2cm}
   \end{minipage}}
  \subfigure[]{
   \label{fig:9b}
   \begin{minipage}[]{0.475\linewidth}
   \centering \includegraphics[width=0.9\linewidth]{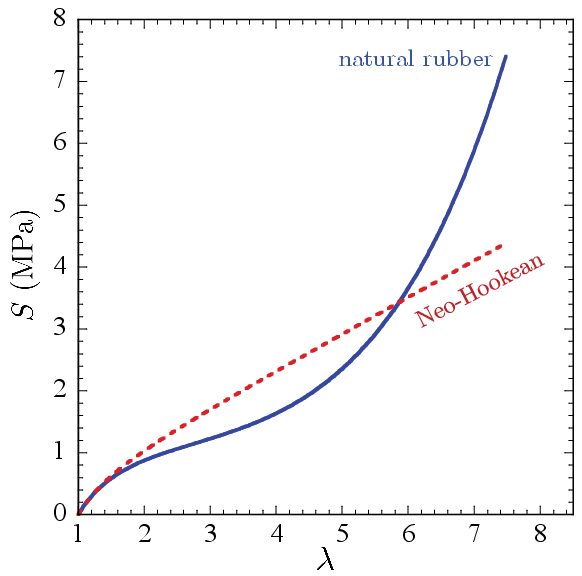}
   \vspace{0.2cm}
   \end{minipage}}
   \caption{(a) Stress-stretch response under uniaxial tension of one of the natural rubbers (vulcanizate \texttt{D}) used by \cite{GL59}, one of the silicone elastomers (PDMS 45:1) used by \cite{Poulain17}, and the SBR used by \cite{Euchler20} in their experimental studies of cavitation. (b) Comparison of the stress-stretch response of the natural rubber with its Neo-Hookean approximation.}\label{Fig9}
\end{figure}

By way of examples, Table \ref{Table1} records the values of the four material constants $\mu_1$, $\alpha_1$, $\mu_2$, $\alpha_2$ in the two-term Lopez-Pamies stored-energy function
\begin{equation}
W=\left\{\begin{array}{ll}\displaystyle\sum_{r=1}^2\frac{3^{1-\alpha_r}\mu_r}{2\alpha_r}\left((\lambda_1^2+\lambda_2^2+\lambda_3^2)^{\alpha_r}-3^{\alpha_r}\right) & {\rm if}\;\lambda_1 \lambda_2\lambda_3=1 \vspace{0.1cm}\\
+\infty & {\rm else}\end{array}\right.\label{W-LP-2-term}
\end{equation}
obtained by fitting the experimentally measured response from tension tests\footnote{Unlike \cite{Poulain17} and \cite{Euchler20}, \cite{GL59} did not report direct data for uniaxial tension, but they did report data for the normalized global force-deformation ($S$ \emph{vs}. $\lambda$) response of a thick poker-chip specimen with diameter-to-thickness ratio $D/H=2$ that only deformed without exhibiting cavitation. The elastic response of the natural rubber that they labeled vulcanizate \texttt{D} is extracted from simulating that poker-chip test \citep{KLP21}.} of one of the natural rubbers (vulcanizate \texttt{D}) used by \cite{GL59}, one of the silicone elastomers (PDMS 45:1) used by \cite{Poulain17}, and the SBR used by \cite{Euchler20} in their experiments on cavitation. Figure \ref{Fig9}(a) plots the corresponding response of each of these elastomers under uniaxial tension in terms of the nominal stress $S={\rm d}W(\lambda,\lambda^{-1/2},\lambda^{-1/2})/{\rm d}\lambda$ as a function of the applied stretch $\lambda$, from $\lambda=1$ up to the final stretch reported experimentally. Qualitatively, the non-Gaussian stiffening is obvious from the figure. To further illustrate this non-Gaussian behavior, Fig. \ref{Fig9}(b) compares the response of the natural rubber with that of its corresponding Neo-Hookean approximation (with the same initial shear modulus $\mu=\mu_1+\mu_2=0.588$ MPa). Quantitatively, the values of the material constant $\alpha_2$ in Table \ref{Table1} show that the power $p$ in the growth condition (\ref{Growth-3}) are roughly $p=\alpha_2=2.725, 2.050, 2.150>3/2$ and hence fall squarely within the typical range (\ref{p growth}). Similar results are obtained by using any other non-Gaussian elasticity model.

%
%

Given that the elasticity of the above three elastomers satisfies the growth condition (\ref{Growth-3}) with (\ref{p growth}) and hence (\ref{Growth}), infinitesimally small defects in these materials cannot grow to a finite size by elastic deformation.

More generally, as the following example demonstrates, direct simulations show that small but finite-size defects in these materials remain in the order of their initially small size, irrespective of the applied finite stretch. In particular, if the defects are initially submicron in size, they remain submicron in size --- and hence not optically visible --- irrespective of the applied finite stretch.

Yet, the sudden appearance of optically visible internal cracks was indeed observed in the experiments reported by \cite{GL59}, \cite{Poulain17}, and \cite{Euchler20}. \emph{Ergo}, the experimental observations of cavitation reported in these works --- which are representative of experimental observations of cavitation at large --- cannot be the result of defects that just grow elastically.

\begin{figure}[t!]
\centering
\centering\includegraphics[width=0.95\linewidth]{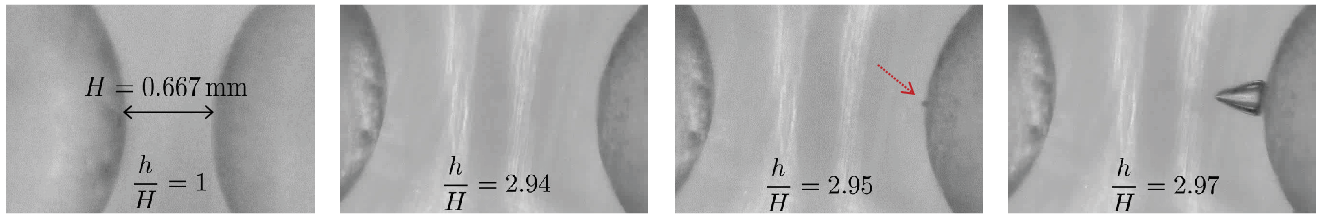}
\caption{\small Selected sequence of images of the two-particle experiment of \cite{Poulain17} on PDMS Sylgard 184 with a 45:1 weight ratio of base elastomer to cross-linking agent. The images show that the first sudden appearance of an optically visible crack occurs 20 $\mu$m away from the inner pole of one of the particles at the global stretch $h/H=2.95$ between the particles; this is indicated by an arrow. Further loading leads to the growth of that crack (shown by the image at $h/H=2.97$) and eventually to the nucleation of another crack (not shown here).}\label{Fig10}
\end{figure}
\begin{figure}[b!]
\centering
\centering\includegraphics[width=0.93\linewidth]{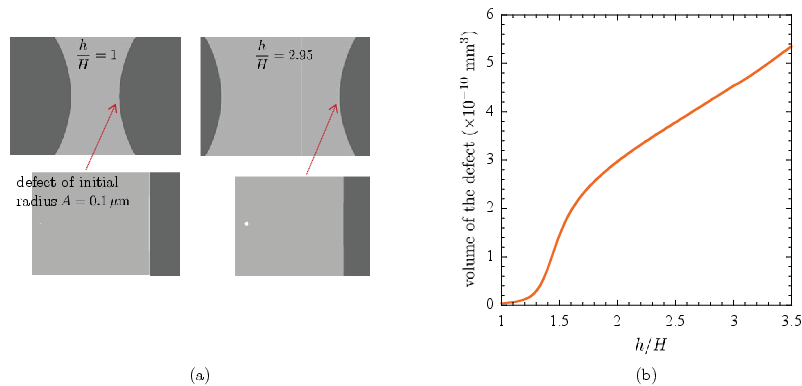}
\caption{\small Simulation of the experiment presented in Fig. \ref{Fig10}. (a) Images showing that a defect, modeled as a vacuous spherical cavity of radius $A=0.1$ $\mu$m, placed 20 $\mu$m away from the inner pole of one of the particles --- the location where cavitation first occurs in the experiment --- exhibits very limited growth at the global stretch $h/H=2.95$ at which cavitation first occurs in the experiment. (b) The volume of the defect as a function of the global stretch $h/H$. Note that even at the global stretch $h/H=3.5$ --- well after cavitation has occurred in the experiment --- the volume of the defect remains below $6\times 10^{-10}$ mm$^3$ and hence not visible under an optical microscope.}\label{Fig11}
\end{figure}

The example corresponds to a simulation presented by \cite{Poulain17} of one their own two-particle experiments, that of two glass spherical particles of diameter $D=3.178$ mm separated by a initial distance $H=0.667$ mm, so that $H/D=0.210$, embedded in PDMS Sylgard 184 with a 45:1 weight ratio of base elastomer to cross-linking agent. In that experiment, as reproduced in Fig. \ref{Fig10}, the first event of cavitation corresponds to the appearance of a single crack of about 10 $\mu$m in size that is located about 20 $\mu$m away from the inner pole of one of the particles, along the axis of symmetry, when the global stretch between the particles reaches the value $h/H=2.95$.

Figure \ref{Fig11} presents results from the elastic simulation of the experimental results presented in Fig. \ref{Fig10}. In the simulation, the elastic behavior of the elastomer is modeled with the non-Gaussian stored-energy function (\ref{W-LP-2-term}) and the material constants listed in Table \ref{Table1} for silicone, while the glass particles are modeled as essentially rigid solids. A vacuous spherical cavity of radius $A=0.1$ $\mu$m is placed 20 $\mu$m away from the inner pole of one of the particles, along the axis of symmetry. This is done because it is known from the images in Fig. \ref{Fig10} that: ($i$) whatever defects exist in the elastomer, they are not visible under an optical microscope and hence must be necessarily submicron in size, and ($ii$) the first cavitation event occurs roughly 20 $\mu$m away from the inner pole of one of the particles along the axis of symmetry.

The main observation from Fig. \ref{Fig11} is that, precisely because of the non-Gaussian elasticity of the elastomer, the defect remains submicron in size for all the global stretches $h/H\in[1,3.5]$ that are considered in the simulation. As a result, the defect is \emph{never} visible under an optical microscope during the loading process. The fact that an optically visible crack does suddenly appear in the experiment, therefore, cannot be attributed to the elastic growth of a defect.

\subsubsection{A comment on a misleading coincidence}\label{Sec: Comment}

As alluded to at the beginning of this subsection, well-established knowledge that the elasticity of elastomers is non-Gaussian predates the study of cavitation. Yet,  the majority of analyses of cavitation that one finds in the literature, even until this day, ignore this fact and, instead, make use of Neo-Hookean elasticity. The origin of such a widespread tendency to make use of an incorrect assumption can be traced back to an interesting albeit misleading finding of \cite{GL59}.

\begin{figure}[H]
\centering
\centering\includegraphics[width=0.425\linewidth]{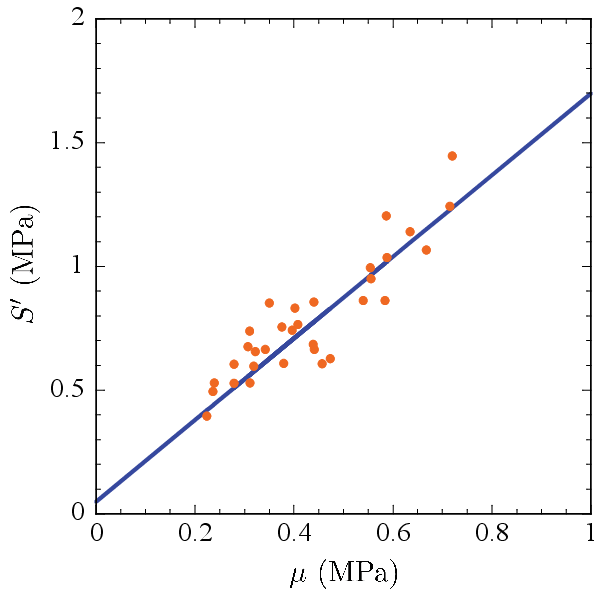}
\caption{\small Relation between the ``local-maximum force'' $S^{\prime}$  and the initial shear modulus $\mu$ noted by \cite{GL59} from poker-chip experiments ($H=0.3$ cm) on different unfilled elastomers; cf. Fig. 7 in their paper.}\label{Fig12}
\end{figure}

\cite{GL59} noted from their poker-chip experiments that the value $S^{\prime}$ of the normalized global force $S=4P/(\pi D^2)$ at the first local maximum in the global force-deformation response --- a presumed proxy for the onset of cavitation --- of specimens with the same initial thickness $H$ depended approximately linearly on the initial modulus $\mu$ (truly, in their analysis, the Young's modulus $E=3\mu$) of the elastomers that they investigated. Figure \ref{Fig12} reproduces their observation for specimens with initial thickness $H=0.3$ cm based on the results from the different types of unfilled elastomers that they tested.

It was the data in Fig. \ref{Fig12} that misled \cite{GL59} into thinking that their experimental observations of cavitation were directly related to the initial shear modulus $\mu$ of the elastomers, and hence that the use of Neo-Hookean elasticity --- wherein the only material constant is $\mu$ --- was justified. But such an interpretation is based on the logical fallacy
$$\emph{cum hoc ergo propter hoc}.$$
Gent himself recognized this much during his Charles Goodyear Medal address \citep{Gent90}, when he noted that
$$\emph{``the fact that a theory appears to work does not mean that it is true''}$$
in reference to the original criterion (\ref{CC-GL}). Indeed, the data in Fig. \ref{Fig12} pertains to four different types of natural rubbers, with four different cross-link densities, and one SBR. Their elasticity is \emph{not} Neo-Hookean but non-Gaussian with strong growth conditions and consequently, as detailed above, does \emph{not} allow for the elastic growth of small defects to finite size.

The fact that Fig. \ref{Fig12} shows an approximately linear relation between the critical normalized global force $S^{\prime}$ and the initial shear modulus $\mu$ of the five unfilled elastomers tested by \cite{GL59} is just a coincidence. As noted by \cite{KLP21}, this coincidence is not entirely surprising because having different cross-link densities does not only imply that the shear moduli of these elastomers are different in a certain manner (i.e., higher cross-link density implies larger $\mu$), but also that other mechanical properties --- for instance, the strength --- may be different in a similar manner as well.

\subsection{The location where cavitation occurs does \emph{not} correlate with the location where defects would grow elastically}\label{Sec: Location}

Even if the elasticity of elastomers is chosen --- contrary to experimental evidence --- to be characterized by a stored-energy function $W$ that does not satisfy the growth condition (\ref{Growth}) and hence allows for the growth of defects to finite size --- such as the popular Neo-Hookean stored-energy function (\ref{W-NH}) --- the experimental observations of cavitation still cannot be explained by the elasticity view. This is because the location of cavitation that elasticity predicts, in general, does not agree with the location where cavitation occurs in experiments.

This discrepancy is most clearly illustrated by the poker-chip experiment. Indeed, the simulations of poker-chip experiments carried out by \cite{LRLP15} for elastomers with Neo-Hookean elasticity have shown that, for sufficiently thin specimens with diameter-to-thickness ratios  $D/H>10$, the first defect that grows elastically to an optically visible size is \emph{always} the one at the center of the specimen\footnote{The same result is true for elastomers with other (\emph{not} Neo-Hookean) elastic behaviors that do not satisfy the growth condition (\ref{Growth}) and hence allow for the growth of defects to finite size.} (assuming, of course, that the distribution of defects is statistically uniform throughout the specimen). As recalled in Subsection \ref{Sec:Beyond Infinitesimal} above from the analysis of \cite{LRLP15}, this is consistent with the fact the criticality condition (\ref{IH-Neo}) is first violated around the centerline of the specimen due to the large hydrostatic stress that develops there.

Yet, as noted in Subsection \ref{Sec: Poker-Chip} above, although the founding poker-chip experiments of \cite{GL59} on natural rubber (on the transparent vulcanizate \texttt{G} in particular) showed that the first optically visible cracks did appear around the center of thin specimens, the poker-chip experiments of \cite{Euchler20} on SBR and those of \cite{GuoRavi23} on silicone elastomers have shown that the location of the first nucleated cracks can be significantly away from the specimen centerline.

\begin{figure}[H]
\centering
\centering\includegraphics[width=0.98\linewidth]{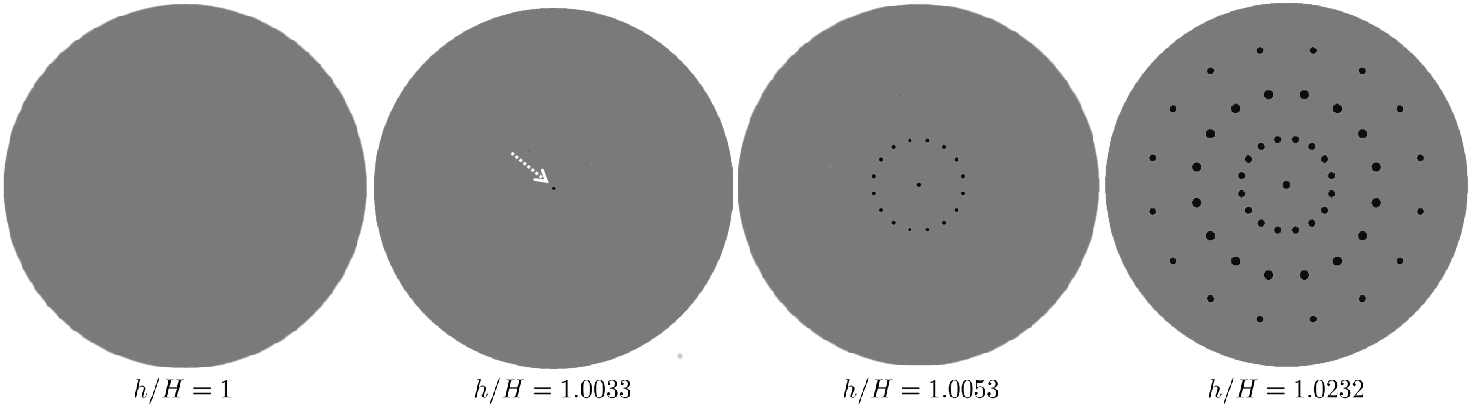}
\caption{\small Simulation by \cite{LRLP15} of a poker-chip experiment for a specimen with diameter $D=2$ cm and initial thickness $H=0.056$ cm, and hence diameter-to-thickness ratio $D/H=35.7$, made of an elastomer with Neo-Hookean elasticity. The images show the sequence in which the defects in the midplane of the specimen grow to a large size as the global stretch $h/H$ between the fixtures increases. Crucially, the defect that grows first is the one at the center of the specimen; this is indicated by an arrow.}\label{Fig13}
\end{figure}

By way of an example, Fig. \ref{Fig13} shows results from the simulation by \cite{LRLP15} of a poker-chip experiment with specimen diameter $D=2$ cm and initial thickness $H=0.056$ cm, so that $D/H=35.7$. The specimen contains a total of 195 defects, which are modeled as vacuous spherical cavities of radius $A=1$ $\mu$m; see Fig. \ref{Fig8}(a) for their FE discretization. The results in Fig. \ref{Fig13} are images of the midplane of the specimen at select values of the applied global stretch $h/H$ between the fixtures. They show that the first defect to grow to a large size is the one at the center of the specimen when the global stretch between the fixtures reaches the value $h/H=1.0033$. Further stretch leads to the growth of adjacent midplane cavities in a radial cascading sequence.

\begin{figure}[t!]
\centering
\centering\includegraphics[width=0.8\linewidth]{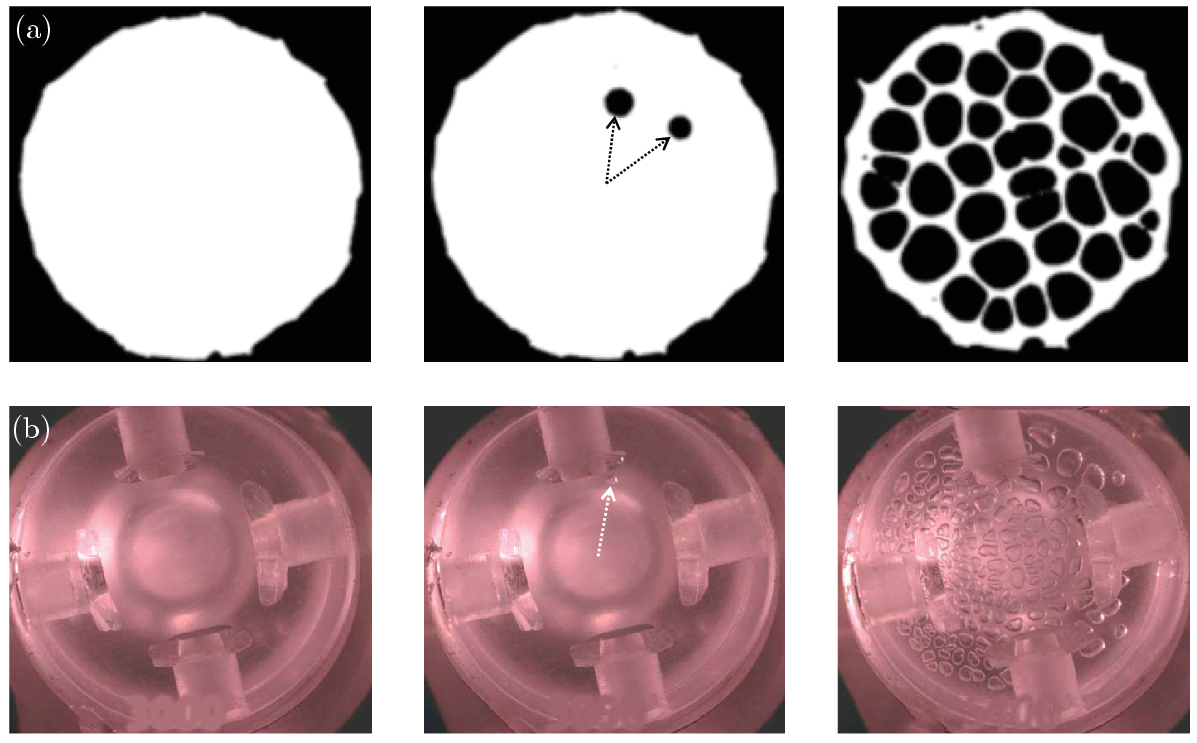}
\caption{\small Selected sequence of images of the poker-chip experiments (a) of \cite{Euchler20} on SBR for a thin specimen with diameter-to-thickness ratio $D/H=20$ and (b) of \cite{GuoRavi23} on PDMS Sylgard 184, with a 30:1 weight ratio of base elastomer to cross-linking agent, for a thinner specimen with diameter-to-thickness ratio $D/H=41.6$. The images show that the first sudden appearance of optically visible cracks (indicated by arrows) occurs significantly away from the centerline of the specimens. Further loading leads to the nucleation of many cracks.}\label{Fig14}
\end{figure}

Figure \ref{Fig14} shows results from the poker-chip experiments of \cite{Euchler20} and  \cite{GuoRavi23}. The experiment of \cite{Euchler20} shown in part (a) of the figure is that of a specimen with diameter $D=1$ cm and initial thickness $H=0.05$ cm, so that $D/H=20$, wherein the elastomer is the SBR with the uniaxial-tension response shown in Fig. \ref{Fig9}. The experiment of \cite{GuoRavi23} shown in part (b), on the other hand, pertains to a specimen with diameter $D=2.5$ cm and initial thickness $H=0.06$ cm, so that $D/H=41.6$, wherein the elastomer is PDMS Sylgard 184 with a 30:1 weight ratio of base elastomer to cross-linking agent. In both experiments, the first event of cavitation corresponds to the appearance of cracks \emph{not} at the center but about $D/4$  away from the centerline of the specimens. Such a location cannot be explained by the elastic growth of defects.

\section{Final comments}\label{Sec:Final comments}

\subsection{The top-down or macroscopic fracture view of cavitation}

As reviewed above, the experimental observations of cavitation are not consistent with the elastic deformation of defects in elastomers.

Having ruled out elasticity, it follows that cavitation is necessarily an inelastic phenomenon. Since, once more, the observations correspond to the sudden appearance of internal cracks, it is natural to consider that the particular type of inelasticity is the creation of new surface. In other words, it is natural to consider that cavitation is nothing more than the nucleation of fracture within the bulk of elastomers.

From a continuum point of view, there are essentially two different approaches to describe nucleation of fracture within the bulk of any (\emph{not} just elastomers) material: ($i$) the bottom-up or microscopic approach and ($ii$) the top-down or macroscopic approach. The microscopic approach is based on the explicit modeling of the defects inherent to the material and of how the new surface that is created from those defects grows to reach the macroscopic length scale at which cracks are observed experimentally. On the other hand, the macroscopic approach consists in modeling the pertinent critical macroscopic fields --- for instance, the critical stresses and energies --- at which the cracks that are observed experimentally nucleate. This is done in an initially ``perfect'' material, since the defects are not modeled explicitly in this approach, instead, it is their macroscopic manifestation that is modeled.

At present, because of the difficulty in generating direct experimental knowledge of the specifics of material defects (this is especially true for submicron defects in elastomers) and of how new surface is created from them, the microscopic approach continues to remain out of reach; insomuch that the state of the art still essentially amounts to the elementary analysis of \cite{WS65}.

By contrast, thanks to the availability of numerous diagnostic tools that can provide direct measurements at length scales of micrometers and larger and at time scales of milliseconds and larger, the macroscopic approach can be pursued with full confidence that it can be based on direct experimental observations.

In this context, based on the experimental findings on cavitation summarized in Section \ref{Sec: Experiments} above and the plethora of experimental results accumulated in the literature since the early 1900s on elastomers with pre-existing cracks, notches, and propagating cracks \citep{Busse34,RT53,Greensmith55,Thomas94,Hamed2016,Chen17}, \citet*{KFLP18} and  \cite{KLP20} have shown that there are three basic ingredients that any attempt at a complete macroscopic description of nucleation and propagation of fracture in elastomers must account for:
\begin{enumerate}[label=\Roman*.]

\item{the stored-energy function $W(\bfF)$ describing the elasticity of the elastomer for arbitrary deformation gradients $\bfF$,}

\item{the strength surface $\mathcal{F}(\bfT)=0$ describing the strength of the elastomer for arbitrary stress tensors $\bfT$, and}

\item{the critical energy release rate $G_c$ describing the intrinsic fracture energy of the elastomer, that is, the amount of energy per unit undeformed area required to create new surface in the elastomer.}

\end{enumerate}
This is because, when viscous dissipation is negligible (more on this in Subsection \ref{Sec: Viscous} below), the experimental observations have shown that \emph{nucleation of fracture}
\begin{itemize}

\item{in a body under a uniform state of stress is governed by the strength of the elastomer,}

\item{from large\footnote{``Large'' refers to large relative to the characteristic size of the underlying heterogeneities in the elastomer under investigation. By the same token, ``small'' refers to sizes that are of the same order or just moderately larger than the sizes of the heterogeneities.} pre-existing cracks is governed by the Griffith competition between the elastic energy and the intrinsic fracture energy,}

\item{under any other circumstance (e.g., from boundary points, smooth or sharp, small pre-existing cracks, or any other subregion in the body under a non-uniform state of stress) is governed by an interpolating interaction among the strength and the Griffith competition (see, e.g., Fig. \ref{Fig16} below),}

\end{itemize}
while \emph{propagation of fracture}
\begin{itemize}

\item{is, akin to nucleation from large pre-existing cracks, also governed by the Griffith competition between the elastic and fracture energies.}

\end{itemize}
Out of the three basic ingredients I-III, the strength is the one that has been more often misunderstood, or outright forgotten, in the literature. It is also the one that dominates the phenomenon of cavitation. We elaborate on this key point next.

\subsubsection{The definition of strength for elastomers}

When a macroscopic piece of the elastomer of interest is subjected to a state of monotonically increasing \emph{uniform} but otherwise arbitrary stress, fracture will nucleate from one or more of its inherent  defects at a critical value of the applied stress. Assuming that viscous dissipation is negligible (see Subsection \ref{Sec: Viscous} below for the case when viscous dissipation is \emph{not} negligible), the set of all such critical stresses defines a surface in stress space. In terms of the Cauchy stress tensor $\bfT$, we write
\begin{equation}
\mathcal{F}(\bfT)=0\label{F-surf-T}
\end{equation}
and choose, as a sign convention, that any stress state such that
\begin{equation*}
\mathcal{F}(\bfT)\geq0
\end{equation*}
is in violation of the strength of the elastomer. Alternatively, in terms of the first Piola-Kirchhoff stress tensor $\bfS$, we write
\begin{equation}
\mathcal{F}_0(\bfS)=0\quad {\rm and} \quad \mathcal{F}_0(\bfS)\geq0.\label{F-surf-S-viol}
\end{equation}
Of course, the description of strength can be done in any other stress measure of choice, however, not all of them are equally convenient. For elastomers that can undergo very large deformations, the strength surface (\ref{F-surf-S-viol})$_1$ in terms of the first Piola-Kirchhoff stress turns out to be more convenient than the strength surface (\ref{F-surf-T}) in terms of the Cauchy stress \citep{KLP20}.

The following three remarks are in order.

\begin{enumerate}[label=\roman*.]

\item{The strength surface (\ref{F-surf-S-viol})$_1$ is an intrinsic macroscopic material property, one that is potentially very different for different elastomers (see, e.g., Fig. \ref{Fig15} below). As such, much like the initial shear modulus $\mu$, it can be directly measured from experiments.}

\item{The experimental measurement of the entire strength surface (\ref{F-surf-S-viol})$_1$ is difficult. This is because it is extremely challenging to subject a specimen to uniform states of stress spanning all triaxialities.

    Indeed, direct measurements of strength in the literature are narrowly restricted to uniaxial \citep{Smith64} and biaxial \citep{Kawabata73} tensile strengths, $s_{\texttt{ts}}$ and $s_{\texttt{bs}}$, which are generated by subjecting specimens to the uniform stresses $\bfS={\rm diag}(s,0,0)$ and $\bfS={\rm diag}(s,s,0)$ until fracture nucleates in the gauge region.

    It is only recent \citep{KLP20,KLP21} that indirect measurements --- in the sense that they are obtained from specimens that are subjected to non-uniform triaxial stress states throughout --- have been utilized to deduce other points beyond $s_{\texttt{ts}}$ and $s_{\texttt{bs}}$ on the strength surface (\ref{F-surf-S-viol})$_1$. In particular, the focus has been in deducing the hydrostatic strength $s_{\texttt{hs}}$, that is, the point defined by the equation $\mathcal{F}_0({\rm diag}(s_{\texttt{hs}},s_{\texttt{hs}},s_{\texttt{hs}}))=0$.

\begin{figure}[h!]
\centering
\centering\includegraphics[width=0.425\linewidth]{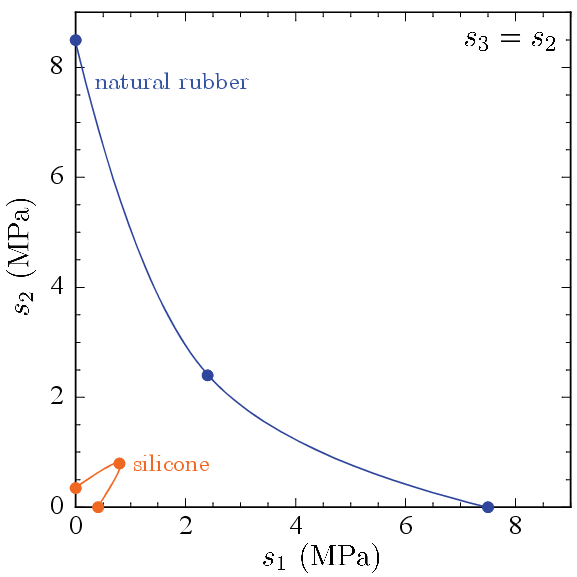}
\caption{\small Strength surface in the space of first Piola-Kirchhoff stresses $\bfS={\rm diag}(s_1,s_2,s_3=s_2)$ of one of the natural rubbers (vulcanizate \texttt{D}) used by  \cite{GL59} and one of the silicone elastomers (PDMS 30:1) used by \cite{GuoRavi23} in their experimental studies of cavitation. Note that while the natural rubber is weakest in hydrostatic tension ($s_2=s_1$), the silicone elastomer is weakest in uniaxial tension ($s_2=0$).}\label{Fig15}
\end{figure}

By way of an example, Fig. \ref{Fig15} presents the strength surface for one of the natural rubbers (vulcanizate \texttt{D}) used by \cite{GL59} and one of the silicone elastomers (PDMS 30:1) used by  \cite{GuoRavi23} in their poker-chip experiments. Specifically, Fig. \ref{Fig15} shows a cross-section of the strength surface (\ref{F-surf-S-viol})$_1$ in the space of axisymmetric first Piola-Kirchhoff stresses $\bfS={\rm diag}(s_1,s_2,s_3=s_2)$. For each elastomer, the figure plots the interpolation (solid lines) between three data points (solid circles): the uniaxial tensile strength $s_{\texttt{ts}}$, the biaxial tensile strength $s_{\texttt{bs}}$, and the hydrostatic strength $s_{\texttt{hs}}$.

Since \cite{GL59} did not report strength data, the values of $s_{\texttt{ts}}$ and $s_{\texttt{bs}}$ plotted in Fig. \ref{Fig15} for the natural rubber correspond to the experimental values measured by other researchers \citep{Smith64,Kawabata73} on a natural rubber with similar cross-link density. The value for $s_{\texttt{hs}}$ corresponds to that extracted by  \cite{KLP21} from the poker-chip experiment of \cite{GL59} on their thinnest specimen, the one with initial thickness $H=0.056$ cm, where the stress field was closest to uniform; see Section 4.2 in \citep{KLP21} and Fig. \ref{Fig16} below. On the other hand, the value of $s_{\texttt{ts}}$ plotted in Fig. \ref{Fig15} for the silicone elastomer corresponds to the experimental value measured by  \cite{Poulain17}, while the value $s_{\texttt{hs}}$ corresponds to that extracted by \cite{KLP20} from a two-particle experiment of \cite{Poulain18}, and that of  $s_{\texttt{bs}}$ is an extrapolation of $s_{\texttt{ts}}$ and $s_{\texttt{hs}}$ based on experimental data \citep{Mazza16} for $s_{\texttt{bs}}$ for the same type of silicone elastomer with a higher cross-link density (PDMS 10:1).

The main observation from Fig. \ref{Fig15} is that different elastomers can have very different strength surfaces, not only quantitatively but also qualitatively. Indeed, while the natural rubber exhibits a strength that is weakest in hydrostatic tension, the silicone elastomer features a strength that is weakest in uniaxial tension (and strongest in hydrostatic tension).

}

\item{Physically, the strength surface (\ref{F-surf-S-viol})$_1$ is the macroscopic manifestation of the presence of defects in the elastomer. These can be of different natures. Furthermore, their size and spatial variations are inherently stochastic. These variations are most acute when comparing material points within the bulk of the body with material points on its boundary, since different fabrication processes or boundary treatments can drastically affect the nature of boundary defects vis-a-vis those in the bulk. It is for this reason that the strength surface (\ref{F-surf-S-viol})$_1$ is an intrinsic material property that is \emph{not} deterministic but inherently stochastic; for clarity of presentation, we have not shown the stochasticity of the strength surfaces in Fig. \ref{Fig15}.}

\end{enumerate}

\subsubsection{The role of the strength in cavitation}\label{Sec: strength liquids}

As anticipated above, out of the three basic ingredients I-III required to formulate a complete description of fracture, the strength surface $\mathcal{F}_0(\bfS)=0$ is the one that dominates the phenomenon of cavitation in elastomers. This is because cavitation occurs within the bulk of elastomers in regions where the stress field is typically only moderately non-uniform and, in particular, not singular.

By definition, as elaborated in the preceding subsection, the attaintment of the strength criticality condition $\mathcal{F}_0(\bfS)=0$ implies nucleation of fracture only when the stress field is uniform. When the stress field is non-uniform, the satisfaction of the strength condition $\mathcal{F}_0(\bfS)\geq0$ at a material point is a necessary but \emph{not} sufficient condition for fracture nucleation to occur at that point. Instead, the nucleation of fracture then is governed neither solely by strength nor solely by the Griffith competition between the elastic and fracture energies, but by an ``interpolation'' between the two \citep{KBFLP20,LP23}.

\begin{figure}[b!]
\centering
\centering\includegraphics[width=0.425\linewidth]{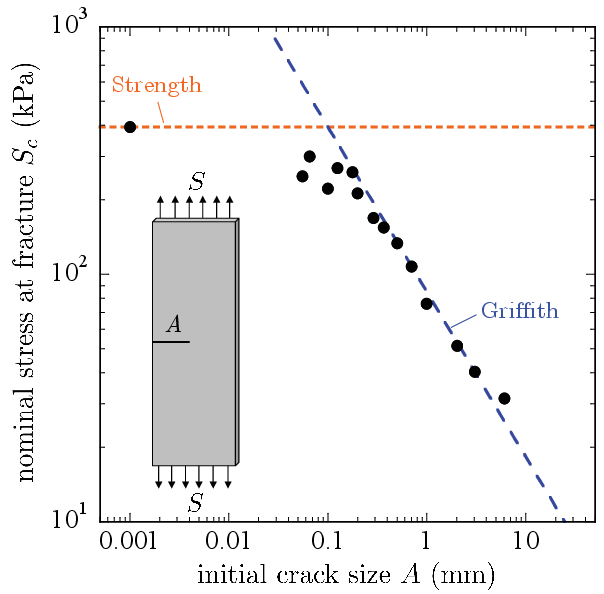}
\caption{\small The experimental data --- from the test schematically depicted in the inset --- of \cite{Chen17} on an acrylic elastomer illustrating the transition from Griffith-dominated to strength-dominated nucleation of fracture as the size of the crack decreases from  $A\approx10$ mm to submicron sizes. The results show the critical nominal stress $S_c$ at which fracture nucleates from the crack as a function of its size $A$. For direct comparison, the plot includes the predictions of nucleation based on the Griffith competition between the elastic and fracture energies (blue dashed line) and based on strength (orange dotted line).}\label{Fig16}
\end{figure}

A classical test (see the inset in Fig. \ref{Fig16}) that illustrates this behavior is that of specimens containing an edge crack of sizes varying from ``small'' to ``large'' that are subjected to a tensile nominal stress perpendicular to the crack. Figure \ref{Fig16} reproduces the results (solid circles) for an acrylic elastomer (VHB 4905 from the 3M Company) obtained by \cite{Chen17} from such a test, wherein the size of the crack was varied from  $A\approx10$ mm to the natural submicron sizes that result on the boundary of the specimens from the fabrication process. The results show that for crack sizes  $A>0.3$ mm, nucleation of fracture is characterized by the Griffith competition between the elastic and fracture energies of the elastomer (blue dashed line). On the other hand, for crack sizes  $A<1$ $\mu$m, nucleation of fracture is characterized by the strength of the material (orange dotted line), in this simple case, its uniaxial tensile strength $s_{\texttt{ts}}\approx 400$ kPa.  Finally, for crack sizes in the intermediate range  $A\in[0.001,0.3]$ mm, the results  show that nucleation of fracture is indeed characterized by an interpolation between the strength and the Griffith competition between the elastic and fracture energies.

Roughly speaking, whether the nucleation of fracture in a given boundary-value problem is dominated more by the strength or by the Griffith competition between the elastic and fracture energies depends on the level of non-uniformity of the stress field. If the stress field varies in space slowly, then the nucleation of fracture will be dominated by the strength and, viceversa, if the spatial gradient of the stress field is large, then the nucleation of fracture will be dominated by the Griffith competition.

\begin{figure}[t!]
\centering
\centering\includegraphics[width=0.80\linewidth]{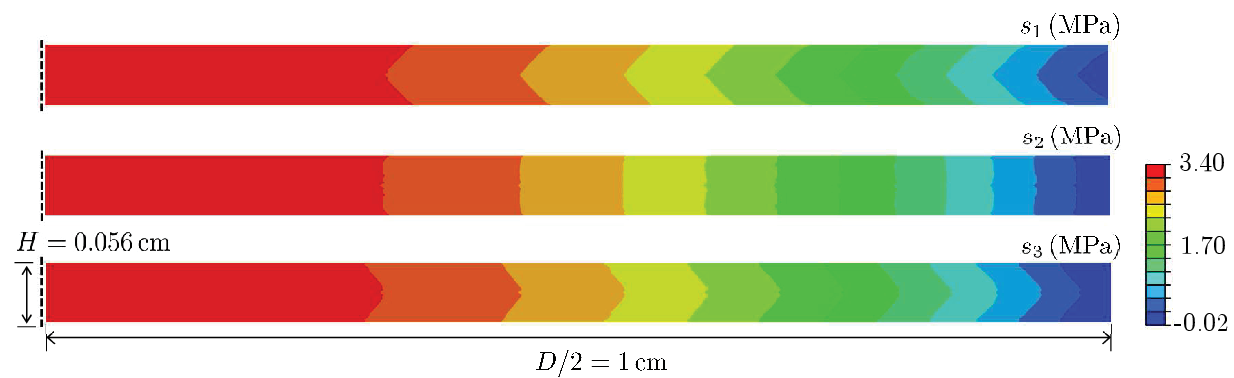}
\caption{\small Simulation by \cite{KLP21} of one of the poker-chip experiments carried out by \cite{GL59} on a specimen with initial thickness $H=0.056$ cm. The images show contour plots over the undeformed configuration of the three principal nominal stresses $s_1$, $s_2$, $s_3$ over half of the specimen (for better visualization) at the normalized global force $S=4P/(\pi D^2)=1.80$ MPa, just before cavitation occurs.}\label{Fig17}
\end{figure}

Accordingly, because the non-uniformity of the stress field is not exceedingly large in regions where cavitation most often occurs, the expectation is that cavitation is dominated by the strength of the elastomer. To illustrate this point, Fig. \ref{Fig17} reproduces from the simulations of \cite{KLP21} the contour plots of the principal nominal stresses $s_1$, $s_2$, $s_3$ over the undeformed configuration of one of the poker-chip specimens tested by \cite{GL59} on vulcanizate \texttt{D}, that with initial thickness $H=0.056$ cm. In the simulation, the elastic behavior of the natural rubber is modeled with a nearly incompressible version of the non-Gaussian stored-energy function (\ref{W-LP-2-term}) and the material constants listed in Table \ref{Table1} for natural rubber. The contours are shown at the normalized global force $S=4P/(\pi D^2)=1.80$ MPa (the corresponding applied deformation is $h/H=1.011$), which is thought to roughly correspond to the instance of the first cavitation event in the experiment.

As expected from the near incompressibility of the elastomer, the first Piola–Kirchhoff stress $\bfS$ throughout a central region of the specimen --- where the nucleation of cracks is believed to first occur --- is roughly uniform and purely hydrostatic. Quantitatively, that central region extends about $D/3$ in radial distance from the centerline and is subjected to a stress field that is roughly given by $\bfS\approx{\rm diag}(3.4,3.4,3.4)$ MPa. At larger distances away from the centerline, the stress field is clearly non-uniform. Nevertheless, the length scale of the stress gradient is about the initial thickness of the specimen $H=0.056$ cm and hence not exceedingly small. In view of this moderate non-uniformity of the stress field, it is reasonable to expect that in this type of experiment cracks will nucleate in regions where the strength surface of the material is violated, soon after is violated.

Now, to establish the verity of this expectation, a complete theory of fracture is needed that accounts for all three basic ingredients I-III  and that interpolates appropriately between the strength and the Griffith competition between the elastic and fracture energies so as to describe fracture nucleation in regions where the stress is non-uniform. In 2018, \citet*{KFLP18} put forth a formulation for one such theory. As already mentioned in the Introduction, direct comparisons \citep{KFLP18,KRLP18,KLP20,KLP21,KKLP23} with both new and classical experiments have shown that their formulation is capable of accurately describing the entire range of experimental observations of cavitation found in the literature. These comparisons appear to confirm the expectation that the phenomenon of cavitation is indeed dominated by the strength of the elastomer.

%
%

\subsubsection{The fracture view vs. experimental observations: A few representative results}\label{Sec:Fracture Comparisons}

In the sequel, for illustration purposes, we reproduce some of the comparisons between the macroscopic fracture theory of \citet*{KFLP18} and experimental observations.

\begin{figure}[h!]
\centering
\centering\includegraphics[width=0.95\linewidth]{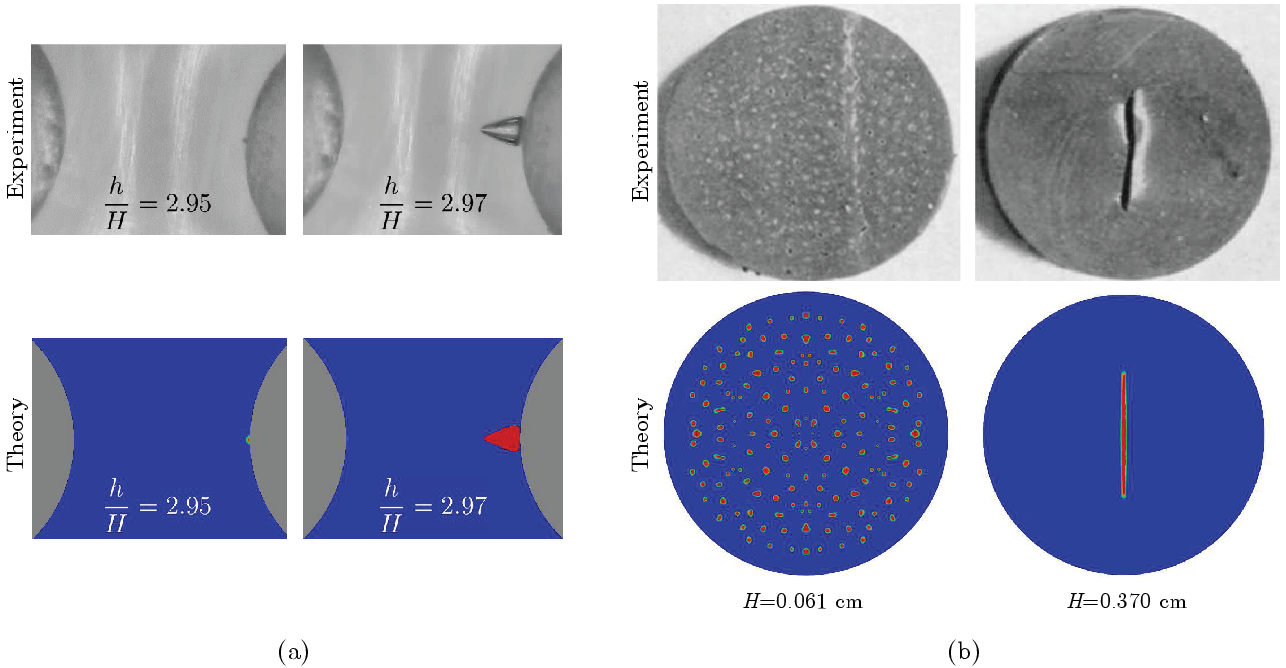}
\caption{\small Representative comparisons between the macroscopic fracture theory of \citet*{KFLP18} with experimental observations of cavitation for (a) a two-particle experiment of \cite{Poulain17} on a silicone elastomer (PDMS 45:1) and (b) two poker-chip experiments of \cite{GL59} on a natural rubber (vulcanizate \texttt{D}). The images in part (a) show the instances at which a crack first nucleates and its subsequent growth. Part (b) shows the post-mortem images of the midplane of specimens with two different initial thicknesses, $H=0.061$ cm and $H=0.370$ cm.}\label{Fig18}
\end{figure}

As a first representative result, Fig. \ref{Fig18}(a) reproduces \citep{KFLP18} a comparison between the macroscopic fracture theory with a two-particle experiment of \cite{Poulain17} on a silicone elastomer, the one already shown in Fig. \ref{Fig10} above. The images pertain to two snapshots during the loading process, the instance at which a crack first nucleates near the inner pole of one the glass particles and a later instance showing the growth of that nucleated crack. In the simulations, the region in red shows that the nucleation, propagation, and deformation of the crack are all in agreement with the experimental observations. In this case, the nucleation of the crack occurs near the inner pole of one of the particles because the strength surface of the silicone elastomer is violated there. Its subsequent growth is described by the Griffith competition between the elastic and fracture energies of the elastomer.

Figure \ref{Fig18}(b) reproduces \citep{KLP21} a comparison between the macroscopic fracture theory with two poker-chip experiments of \cite{GL59} on natural rubber, the ones on vulcanizate \texttt{D} and specimens with initial thicknesses $H=0.061$ cm and $H=0.370$ cm.  The images show the midplane of the specimens, cut open post-mortem. In the simulations, as in part (a) of the figure, the regions in red denote the cracks that have been nucleated and propagated. The agreement in the number, location, and shape of the cracks predicted by theory with those observed experimentally is evident. In this case too, the nucleation of the cracks occurs because the strength surface of the natural rubber is violated and their subsequent growth is described by the Griffith competition between the elastic and fracture energies of the rubber. The reason why the thin specimen has many cracks, while the thick specimen has only one is due to the competition between nucleation and propagation. For the thin specimen, it is more energetically costly to propagate cracks than for the thick specimen. This last point can be readily corroborated by direct calculations of the relevant energy release rate; see, e.g., \citep{Suo23}.

\subsection{Some open problems}

\subsubsection{Experimental measurements of the strength surface}

At this stage, it is plain that the phenomenon of cavitation in elastomers can be explained as a fracture nucleation process. As such, again, when viscous dissipation is negligible, its description and prediction requires having knowledge of the elasticity of the elastomer, as well as of its strength surface and intrinsic fracture energy. Out of these three material properties, standardized tests exist to measure the elasticity and the intrinsic fracture energy. However, little is known about the strength surface (\ref{F-surf-S-viol}) of elastomers beyond the uniaxial and biaxial tensile strength.

The experimental measurement of points on the strength surface $\mathcal{F}_0(\bfS)=0$ of elastomers --- especially in the first octant $s_1,s_2,s_3>0$ in the space of principal nominal stresses $(s_1,s_2,s_3)$ and with emphasis on its stochasticity --- is hence one of the main open problems that needs to be addressed in order to advance our understanding not only of cavitation but of nucleation of fracture at large.

\subsubsection{Accounting for viscous dissipation}\label{Sec: Viscous}

By now it is well established that all elastomers exhibit viscous dissipation and that, more often than not, such a dissipation is not negligible \citep{Ferry80,deGennes79,Wineman2009,KLP16}. When accounting for viscous dissipation --- similar to the case when elastomers are idealized as elastic solids --- there are still three basic ingredients \citep{SLP23,LP24} that any attempt at a macroscopic description of nucleation and propagation of fracture must account for: ($i$) the viscoelasticity of the elastomer, ($ii$) its strength, and ($iii$) its intrinsic fracture energy. Once again, out of these three material properties the strength is the one that is least known.

Precisely, the majority of the strength data available in the literature pertains to elastomers under uniaxial tension, when
$\bfS(t)={\rm diag}(0,0,s(t))$, and under biaxial tension, when $\bfS(t)={\rm diag}(s(t),s(t),0)$, this only when either the stretch or the stress are applied at various constant rates in time $t$ within a certain range. This data --- which seems to have been overlooked in the recent literature --- is mostly due to an experimental campaign led by T.L. Smith \citep{Smith58,Smith60,Smith63,Smith64,Smith64b,Smith65,Smith69} some 70 years ago. Remarkably, the measurements show that, to a first degree of approximation, the strength of elastomers is independent of the history of the loading and hence that it can be represented as a hypersurface ``just'' in stress-deformation space \cite{LP24}. In terms of the first Piola-Kirchhoff stress tensor $\bfS$ and the deformation gradient tensor $\bfF$, we write
\begin{equation}
\mathcal{F}_0\left(\bfS,\bfF\right)=0.\label{F-SF}
\end{equation}

Experiments that probe whether a hypersurface of the form (\ref{F-SF}) does indeed describe the strength of viscoelastic elastomers under arbitrary time-dependent loadings constitute another main open problem to advance our understanding of cavitation.

\subsubsection{Accounting for the presence of a solvent}

In many applications, elastomers contain a solvent, which can be either a gas or a liquid. The occurrence of cavitation has also been observed \citep{GT69,Yamabe11,Zhao17,Castagnet18,Kulkarni23,Creton23} in these material systems since the pioneering work of \cite{GT69} on transparent natural and synthetic rubbers supersaturated with gaseous argon.

In addition to accounting for the effects of viscous dissipation, accounting for the presence of a diffusion-capable solvent is thus another interesting direction of future research on cavitation.

\subsection{A clarifying remark on a class of emerging experimental techniques}\label{Sec:Related Problems}

We close by making a clarifying remark regarding a class of emerging experimental techniques \citep{Crosby07,Franck18,Cohen19,Crosby20b,VanVliet21} --- dubbed ``cavitation rheology'' --- that have been proposed in the literature as new methods to characterize the mechanical properties of soft materials and soft biological tissues. A review of these techniques has been recently given by Crosby and collaborators \citep{Crosby20}.

The first of these techniques was proposed by \cite{Crosby07}. It was in this work that the term ``cavitation rheology'' was introduced. The technique consists in injecting a pressurized fluid, either a gas or a liquid, at the tip of a syringe needle embedded in the soft material of interest. At some critical instance during the injection process, a cavity that is a few fold larger than the diameter of the needle suddenly appears around the tip of the needle. The idea is that one can make use of a model together with an inverse analysis to deduce some of the mechanical properties of the material from the appearance of such a cavity.

Now, in spite of the similar nomenclature, ``cavitation rheology'' experiments are fundamentally different from the classical cavitation experiments summarized in Section \ref{Sec: Experiments} above on two counts.

First, the cavities that are generated in ``cavitation rheology'' experiments often appear on the \emph{boundary} of the specimen, exactly where the boundary conditions are applied. By contrast, the cracks that are generated in classical cavitation experiments appear in the \emph{interior} of the specimen, far away from where the boundary conditions are applied.

Indeed, consider for instance the experiment with a syringe needle proposed by \cite{Crosby07}. Prior to the injection of the fluid, the needle is embedded in the specimen. This creates a complex boundary between the needle and the material being tested. As the injection process begins, the fluid makes direct contact with the specimen. This direct contact is a boundary condition that leads to the deformation and possibly also the fracture of the material around the injected fluid. The observed cavity is the result of this deformation and fracture. Put another way, this experiment can be thought of as an indentation experiment wherein the indenter is made of a fluid.

Second, the initial length scale of the cavities that are generated in ``cavitation rheology'' experiments is only a few fold larger than the length scale of the diameter of the needle, the pre-cursor of the cavities. Moreover, the length scale of the diameter of the needle is relatively large, specifically, larger than 50 $\mu$m for any standard syringe needle. By contrast, the cracks that are generated in classical cavitation experiments are at the very least two orders of magnitude larger than their pre-cursors, the inherent defects of the elastomer, which are typically submicron in size.

Never mind these two fundamental differences, ``cavitation rheology'' experiments are still experiments where the material being tested does two things: it deforms and it fractures. As such, much like it has been done for the classical cavitation experiments, one can make use of the complete fracture theory of \citet*{KFLP18} to analyze them. This would require the accurate description of the applied boundary conditions, which is not trivial.

\section*{Acknowledgements}

Support for this work by the 3M Company and by the National Science Foundation through the Grant DMS--2308169 is gratefully acknowledged.

\bibliographystyle{elsarticle-harv}
\bibliography{References}

\end{document}